\PassOptionsToPackage{table}{xcolor}
\documentclass{fairmeta}
\usepackage{amsmath}
\usepackage{amssymb}
\usepackage{xspace}
\usepackage[symbol]{footmisc}
\usepackage{enumitem}
\usepackage{pgffor}
\usepackage{wrapfig}
\usepackage{float}
\usepackage{multirow}
\usepackage{subcaption}
\usepackage[textsize=scriptsize]{todonotes}
\usepackage{tabularx}
\usepackage{xcolor}

\usetikzlibrary{decorations.pathreplacing, arrows.meta}

\definecolor{assetgenfullrow}{HTML}{DCEBFF}
\definecolor{assetgennanorow}{HTML}{EAF6E6}

\newcommand{\assetgenpro}{AssetGen\xspace}
\newcommand{\assetgenflash}{AssetGen Flash\xspace}

\newcommand{\method}{AssetGen\xspace}

\newcommand{\z}{\boldsymbol{z}}

\definecolor{tabfirst}{rgb}{1, 0.7, 0.7}
\definecolor{tabsecond}{rgb}{1, 0.85, 0.7}
\definecolor{tabthird}{rgb}{1, 1, 0.7}

\definecolor{metabluelight}{HTML}{CCE0F8}
\definecolor{metabluemedium}{HTML}{6D92BF}

\newcommand{\dimt}{{D_{\text{T}}}}

\newcommand{\img}{\mathbf{I}}

\newcommand{\beps}{\boldsymbol{\epsilon}}

\newcommand{\degree}{$^\circ$}

\makeatletter
\DeclareRobustCommand\onedot{\futurelet\@let@token\@onedot}
\def\@onedot{\ifx\@let@token.\else.\null\fi\xspace}

\makeatother

\def\methodA{Commercial Model A}
\def\methodB{Commercial Model B}
\def\methodC{Commercial Model C}
\def\methodD{Commercial Model D}

\usetikzlibrary{decorations.pathreplacing, arrows.meta}

\titleformat{\paragraph}[runin]{\sffamily\bfseries}{}{0pt}{}[]
\titlespacing*{\paragraph}{0pt}{1em}{0.5em}

\newcommand{\Description}[1]{} 

\title{AssetGen: Deployable 3D Asset Generation at Interactive Speed}

\author[\dagger]{Dilin Wang}
\author[]{Xiaoyu Xiang}
\author[]{Kihyuk Sohn}
\author[]{Tom Monnier}
\author[]{Yu-Ying Yeh}
\author[]{Thu Nguyen-Phuoc}
\author[]{Jiawen Zhang}
\author[]{Yuchen Fan}
\author[]{Antoine Toisoul}
\author[]{Hyunyoung Jung}
\author[]{Prithviraj Dhar}
\author[]{Michael Bunnell}
\author[]{Nikolaos Sarafianos}
\author[]{Chuhang Zou}
\author[]{Roman Shapovalov}
\author[\dagger]{Andrea Vedaldi}
\author[\dagger]{Rakesh Ranjan}

\affiliation[]{Reality Labs, Meta}
\contribution[\dagger]{Project Lead}

\abstract{
While 3D generation is progressing rapidly, recent work has often focused on obtaining high-resolution assets, leaving user experience and deployability as afterthoughts. We present AssetGen, a 3D generator that focuses instead on these two aspects. Given one reference image, in 30 seconds it produces a high-quality mesh with baked normals, a color texture, and a controlled polygon budget suitable for real-time rendering, including mobile use cases. The AssetGen Flash variant further reduces latency to 14 seconds for interactive and agentic creation loops. Our model generates the object geometry with a coarse-to-refine VecSet framework, which implements mesh simplification, cleaning, and normal baking on the GPU, and a fast parallel UV unwrapping. It then generates textures in a multi-view fashion, followed by backprojection and 3D inpainting. Model distillation, kernel optimization, and pipeline parallelization are co-designed to accelerate the system end-to-end. We introduce numerous automated and blind human evaluations and demonstrate competitive visual quality against leading commercial solutions in 30 seconds and preview-quality results in less than 15 seconds. The final result is a system that supports AI-assisted, deployable 3D content creation in interactive workflows.

}

\date{\today}

\begin{document}

\maketitle

\begin{figure}[ht]                                                      
\centering 
\includegraphics[width=\textwidth]{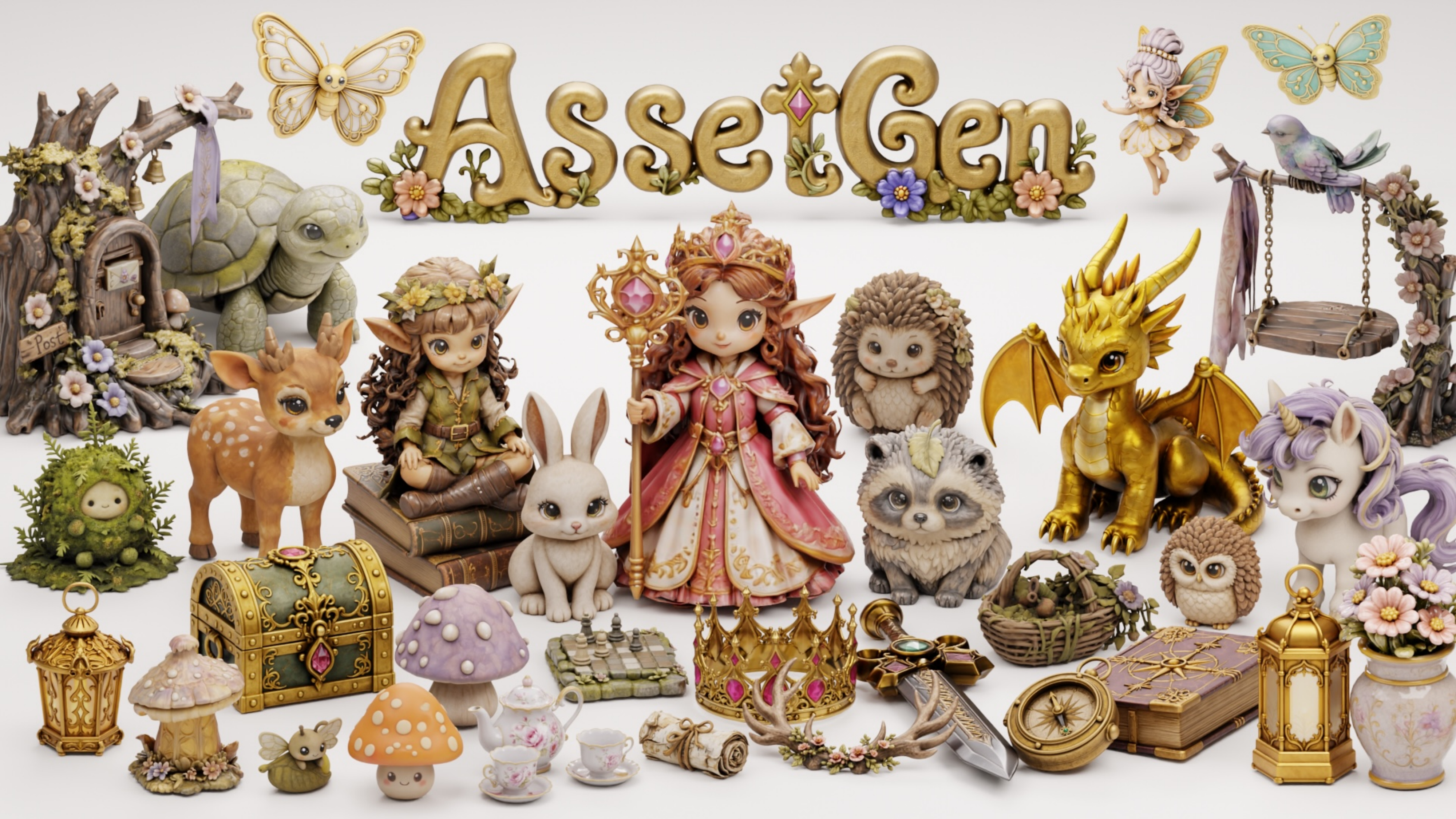}
\vspace{-5mm}
\caption{\textbf{\method.} Our system generates high-fidelity, UV-unwrapped, and normal-baked textured 3D meshes optimized for real-time rendering, delivering production-ready assets in approximately 30 seconds on H100 GPUs.}
\label{fig:teaser}
\Description{A teaser montage showing AssetGen outputs as textured 3D assets optimized for real-time rendering from a single reference image.}
\end{figure}   
 
\tableofcontents

\section{Introduction}%
\label{sec:intro}

3D assets are a key part of games, AR/VR, robotic simulation, and more.
Yet, producing them remains a slow, manual process: artists must shape the geometry of the assets, optimize the mesh and polygon count to stay within a certain budget, unwrap the mesh for texturing, bake normals, and author the texture.
The resulting assets can be excellent, but the workflow is slow and difficult to scale to the volume of content needed by existing and future applications.

Image-to-3D generation has the potential to change the economics of 3D content creation.
In particular, recent work has made significant progress on the quality of generated 3D assets.
Geometry models have improved shape fidelity and detail~\citep{li2025triposg,zhao2025hunyuan3d,xiang2024trellis,xiao2025flashvdm}, while multi-view texture models have improved artistic quality and view consistency~\citep{bensadoun2024meta,cheng2025mvpaint}.
However, while the generated assets may look good at a glance, they are often too complex to use in real-time applications, particularly on mobile devices.
Generation speed can also frustrate users.
A useful generator must have a latency in the tens of seconds to support workflows where assets are proposed, assessed, and revised iteratively by creators.

We therefore consider the problem of generating \emph{application-ready 3D asset} in \emph{interactive creative workflows}.
To be deployable, the assets must satisfy quality and efficiency bars sufficient for use in real-time applications, including mobile ones.
To be useful in interactive workflows, the assets must be generated in seconds rather than minutes.

To this end, we introduce \textbf{\method}, an image-to-3D system designed to meet these goals. As shown in~\cref{fig:teaser}, \assetgenpro delivers a high-quality textured mesh with a controlled polygon budget, UV coordinates, and baked normals in approximately 30 seconds on H100 GPUs\@.
A fast variant, \assetgenflash, reduces latency to approximately 14 seconds to support previews in faster interactive and agentic creation loops.

The system consists of three stages.
First, \textit{MeshGen} predicts a dense, highly detailed 3D shape using a two-stage coarse-to-refine diffusion model.
Second, a GPU-based post-processing stack converts the dense shape into a deployable mesh by simplifying and cleaning the geometry, removing hidden faces, unwrapping UVs, and baking tangent-space normals to preserve high-frequency surface details.
Finally, \textit{TextureGen} synthesizes several views of the textured appearance of the processed mesh and fuses them into a 2048px texture atlas using GPU-based upsampling, backprojection, blending, and 3D-aware inpainting.

A key novelty of \method is that inference is optimized end-to-end.
First, latency is reduced through model distillation in both the geometry and texture generation stages.
For example, diffusion sampling in MeshGen is progressively distilled from 120 denoising steps
down to just 30 steps, 
further doubling speed by also distilling Classifier-Free Guidance (CFG) in a single forward pass.
TextureGen is accelerated in a similar manner.
Several other engineering decisions remove further bottlenecks, including FlashAttention-3, graph compilation, selective low-precision execution, non-blocking transfers, GPU-resident geometry operations, and asynchronous serving, all reducing wall-clock latency without regressing asset quality.
The result is a practical image-to-3D pipeline whose output is competitive with leading commercial systems while running at a latency suitable for creative iteration.

The faster version \assetgenflash skips the MeshGen refinement step, reduces the texture resolution to 1024px omitting upsampling and using more aggressive distillation, including CFG distillation.
While the visual quality is slightly reduced, the much faster generation speed makes interactive creation more fluid and enjoyable.

We validate AssetGen using quantitative benchmarks, qualitative comparisons, and blind human evaluations.
In particular, we introduce new benchmarks and evaluation metrics that measure specific aspects of 3D generation, including, for example, how certain details such as faces and hands are reconstructed in 3D characters---something that was not assessed systematically before to the best of our knowledge.
With this, we show that \method  achieves competitive asset quality against leading commercial systems while substantially reducing latency.
To summarize, our contributions are threefold:

\textbf{Deployable 3D asset generation at interactive rates.}
We refocus image-to-3D as the task of generating 3D assets that can be deployed as-is in real-time rendering engines, including on mobile platforms, while also meeting strict latency constraints to support iterative creative workflows.

\textbf{End-to-end system optimization.}
To achieve these goals, we re-engineer the system as a whole.
We optimize each component of the pipeline, including MeshGen and TextureGen distillation, mesh simplification, hidden-face removal, UV unwrapping, normal baking and texture backprojection, and inpainting.
We introduce system optimizations, including minimizing I/O by running most operations on the same GPU and masking latency via parallelism whenever possible.
We optimize for speed while maintaining the quality of the asset.

\textbf{Asset-level quality-latency Pareto design.}
We introduce two practical configurations: the default \assetgenpro prioritizes visual quality with an end-to-end runtime of approximately 30 seconds, while \assetgenflash prioritizes responsiveness and reaches sub-15-second latency.

\section{Related Work}%
\label{sec:related_work}

\begin{figure*}[t]
\centering
\includegraphics[width=1.0\linewidth]{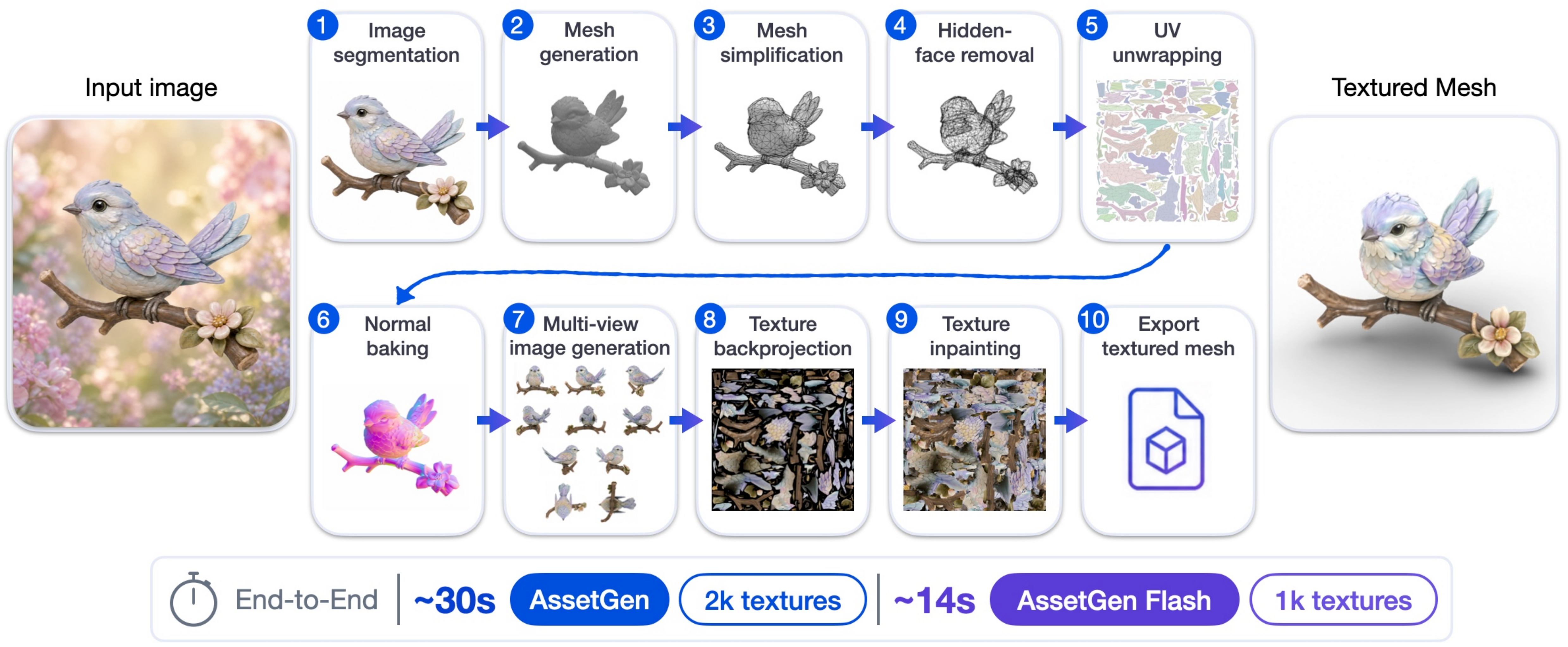}
\vspace{-2mm}
\caption{AssetGen converts a single input image into an export-ready textured mesh with two operating modes: \textbf{AssetGen} denotes the quality-oriented configuration that runs in $\sim 30$ seconds, and \textbf{AssetGen (Flash)} denotes the latency-oriented configuration that runs in approximately $14$ seconds.
Both produce the same asset contract: a simplified mesh with UVs, baked normals, and a texture atlas.}%
\Description{AssetGen pipeline}%
\label{fig:pipeline}
\end{figure*}

\paragraph{3D Shape Generation.} Neural 3D representations trade off rendering quality, extraction cost, and generative scalability.
While radiance fields~\citep{mildenhall2021nerf,muller2022instantngp}, Gaussian splatting~\citep{kerbl2023gaussiansplatting}, and optimization-based lifting~\citep{poole2022dreamfusion,wang2023prolificdreamer} prioritize high-quality view synthesis, interactive pipelines require fast, feed-forward geometric prediction~\citep{TripoSR2024,xu2024instantmesh,wang2024crm,zhou2025spar3d}.
To balance generative scalability with straightforward mesh extraction, recent methods compress 3D data into compact latents like triplanes~\citep{chan2022eg3d}, wavelet features~\citep{wu2024wala}, primitives~\citep{chen2024threedtopia}, structured coordinates~\citep{xiang2025native}, and unordered vector sets~\citep{zhang20233dshape2vecset}.

Building on native 3D diffusion models that learn robust priors directly in these spaces~\citep{zhao2023michelangelo,zhang2024clay,wu2024direct3d,xiang2024trellis,li2025triposg,zhao2025hunyuan3d}, AssetGen adopts a distilled~\citep{xiao2025flashvdm}, two-stage coarse-to-refine VecSet architecture~\citep{lai2025lattice}.
This design provides a compact target for rapid diffusion while supporting the extraction of dense implicit surfaces~\citep{park2019deepsdf,mescheder2019occupancy} necessary for normal baking.
While an emerging family of approaches predicts mesh topology and edge flow directly~\citep{siddiqui2023meshgpt,chen2025deepmesh,meshtron2024,hifimesh2026,xu2026strips,wang2026face,lei2025armesh,lionar2025treemeshgpt,luo2026topomesh,he2025sparseflex}, our system focuses on the pipeline of dense implicit generation followed by explicit processing.
Finally, to ensure high-quality geometry extraction and reliable SDF supervision from curated datasets~\citep{li2025triposg,seed3d2026}, our data pipeline extends established sign-estimation techniques~\citep{newcombe2011kinectfusion,huang2020manifoldplus,barill2018fast} with a boundary-aware flood-fill formulation to handle the open boundaries and inconsistent winding common in real-world assets.

\paragraph{Multi-View Texture Synthesis and Acceleration.} Texture generation has evolved from sequential painting~\citep{richardson2023texture,chen2023text2tex,zeng2024paint3d} to synchronized multi-view synthesis~\citep{liu2023zero,liu2023syncdreamer,shi2023mvdream,long2024wonder3d,voleti24sv3d:,li2024era3d,tang2023mvdiffusion,tang2024mvdiffusion++,cheng2025mvpaint,bensadoun2024meta,mvpainter2025}.
In recent pipelines, these synchronized views are mapped onto explicit surfaces via geometric conditioning and specialized attention mechanisms~\citep{makeatexture2024,sarafianos2025garment3dgen,materialanything2025,videomatgen2026,yang2025hunyuan3d21,he2020epipolar,ye2023ipadapter}.
However, deploying the underlying diffusion transformers~\citep{peebles2023scalable,bao2023all,esser2024scaling,liu2022flow} interactively requires extensive step distillation, guidance optimization~\citep{salimans2022progressive,song2023consistency,sauer2023adversarial,yin2024one,yin2024improved,liu2023instaflow,traflow2026,ho2022classifier,meng2023distillation,cfgzero2026,dice2025}, and kernel-level improvements~\citep{dao2022flashattention,dao2023flashattention2,fa3,torch_compile,qdit2025,diffsparse2026,fast3dcache2025}.
AssetGen co-designs synthesis and acceleration: rather than texturing abstract latents, we condition a synchronized model directly on the deployable, normal-baked mesh.
By combining view-selective attention with distillation and global pipeline optimizations, we achieve high-fidelity texturing within real-time deployment constraints.

\paragraph{Mesh Processing and Production Pipelines.} To satisfy rendering constraints, asset generation relies on foundational geometry processing, including mesh simplification~\citep{garland1997surface,hoppe1996progressive,lindstrom1998fast,decoro2007real}, UV parameterization and seam placement~\citep{levy2002least,sheffer2005abf,young2023xatlas,seamcrafter2025}, and normal baking~\citep{cohen1998appearance}, supported by efficient GPU rendering libraries~\citep{laine2020modular,ravi2020pytorch3d,pidhorskyi2024rasterized,parker2010optix}.
Building on these primitives, complete 3D generation systems increasingly integrate geometry, texture, and export stages~\citep{bensadoun2024meta,zhao2025hunyuan3d,yang2025hunyuan3d21,hunyuan3dstudio2025,boss2025sf3d,xiang2024trellis,xiang2025native,seed3d2026}.
AssetGen contributes a complementary system perspective to this design space: rather than treating model inference, geometry processing, and texture fusion as independent modules, we co-design and globally optimize the full pipeline.
This approach yields an application-ready, UV-unwrapped, and normal-baked mesh while exposing distinct latency-quality operating points for interactive deployment.

\section{System Overview}%
\label{sec:overview}

\method takes a single image as input and generates a textured 3D mesh optimized for real-time rendering.
The system consists of three major stages: geometry generation (MeshGen), geometry processing, and texture generation (TextureGen).
We illustrate this workflow in \cref{fig:pipeline}.
In particular, we use an explicit geometry-first pipeline, in which MeshGen predicts a dense 3D surface that captures global shape and local structure.
The geometry processing stage then converts this dense surface into a lightweight asset through simplification, hidden-face culling, UV unwrapping, and tangent-space normal baking.
TextureGen renders normal and position maps from the processed asset, synthesizes multi-view color images conditioned on those maps and the reference image, and fuses the generated views into the UV atlas through visibility- and incidence-weighted backprojection, with 3D-aware inpainting for regions not observed by any view.
The final output is an explicit textured mesh with a small triangle count, UV coordinates, and a diffuse texture atlas.

We expose two inference configurations with different quality vs. latency trade-offs. The default AssetGen configuration uses two-stage coarse-to-refine MeshGen, runs TextureGen at 1024px per view with inference-time CFG, and applies per-view super-resolution before atlas fusion. This configuration prioritizes visual quality and runs in approximately 30 seconds end to end when deployed on H100 GPUs. AssetGen Flash skips MeshGen refinement, runs TextureGen at 768px per view, uses a guidance-distilled conditional-only TextureGen model, omits per-view super-resolution, and exports a 1K atlas, reducing end-to-end latency to approximately 14 seconds.

Next, we describe data curation (\cref{sec:data_curation}), geometry generation (\cref{sec:meshgen}), geometry post-processing (\cref{sec:geometry_processing}), texture generation (\cref{sec:texturegen}), inference acceleration (\cref{sec:latency-optimization}), evaluation (\cref{sec:evaluation}), qualitative results (\cref{sec:qualitative_results}), conclusion and limitations (\cref{sec:conclusion}).

\section{Data Filtering and Curation}%
\label{sec:data_curation}

\begin{figure}[t]
\centering

\renewcommand{\arraystretch}{0.2}
\setlength{\tabcolsep}{0pt}
\begin{tabular}{cccc}
\includegraphics[width=0.24\linewidth]{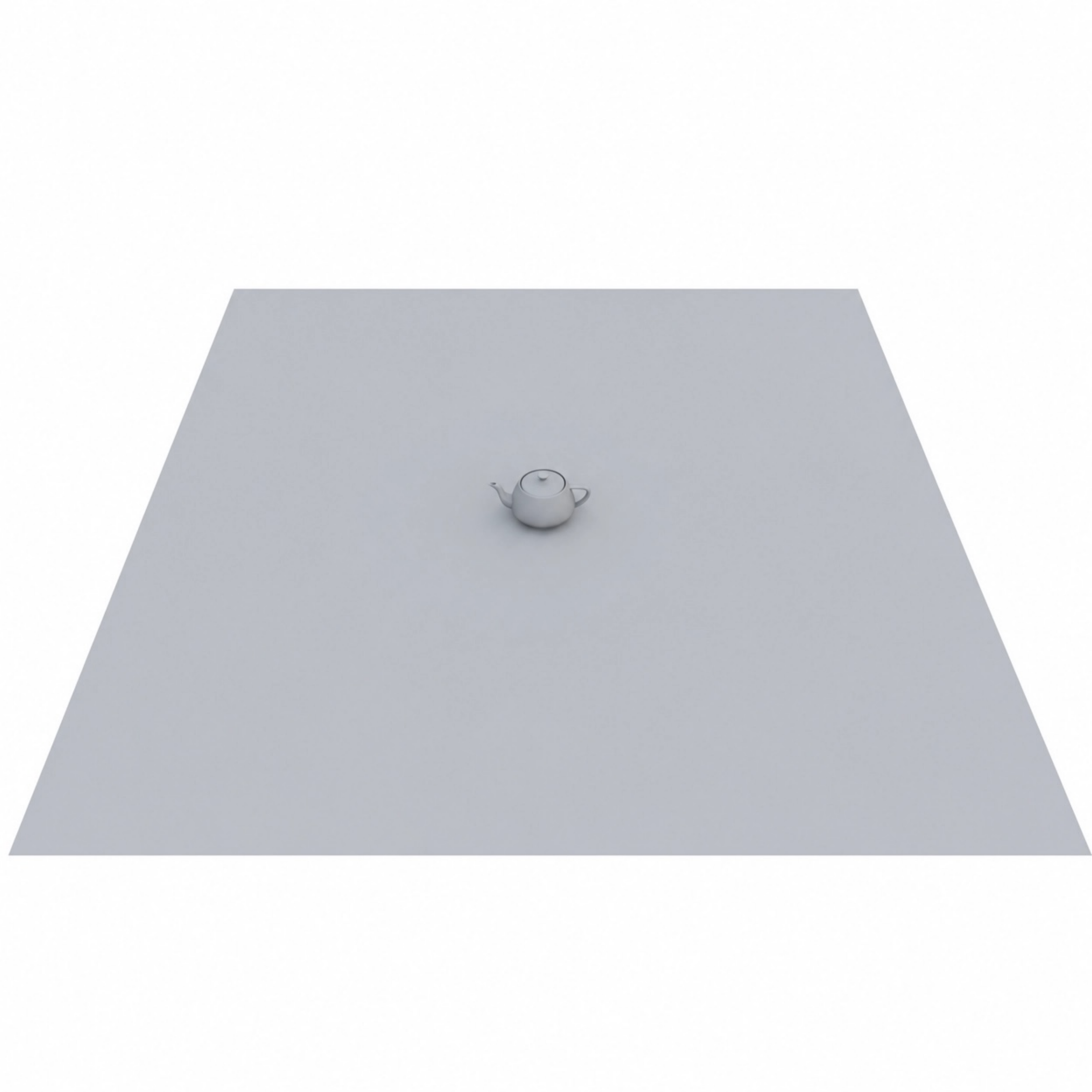} &
\includegraphics[width=0.24\linewidth]{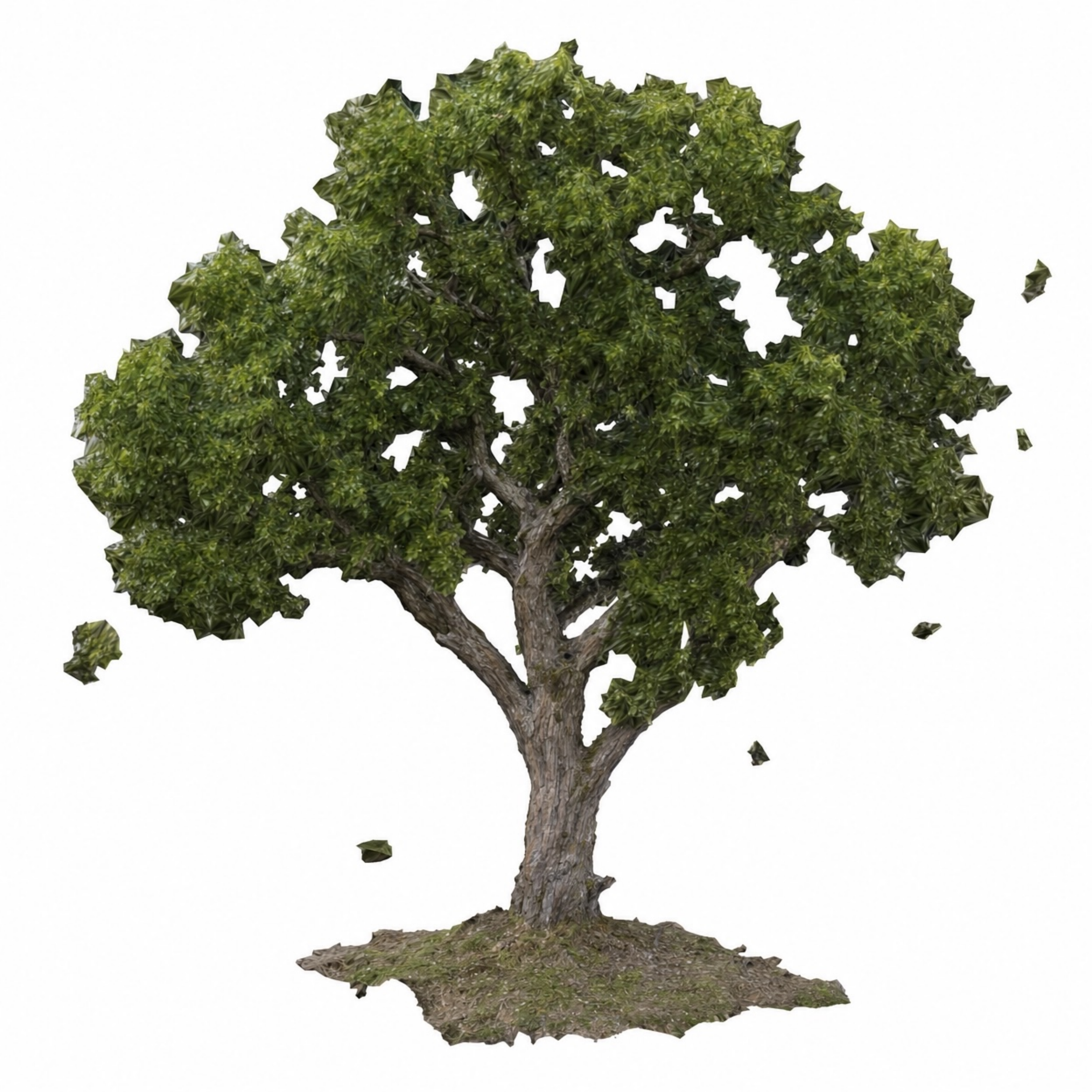} &
\includegraphics[width=0.24\linewidth]{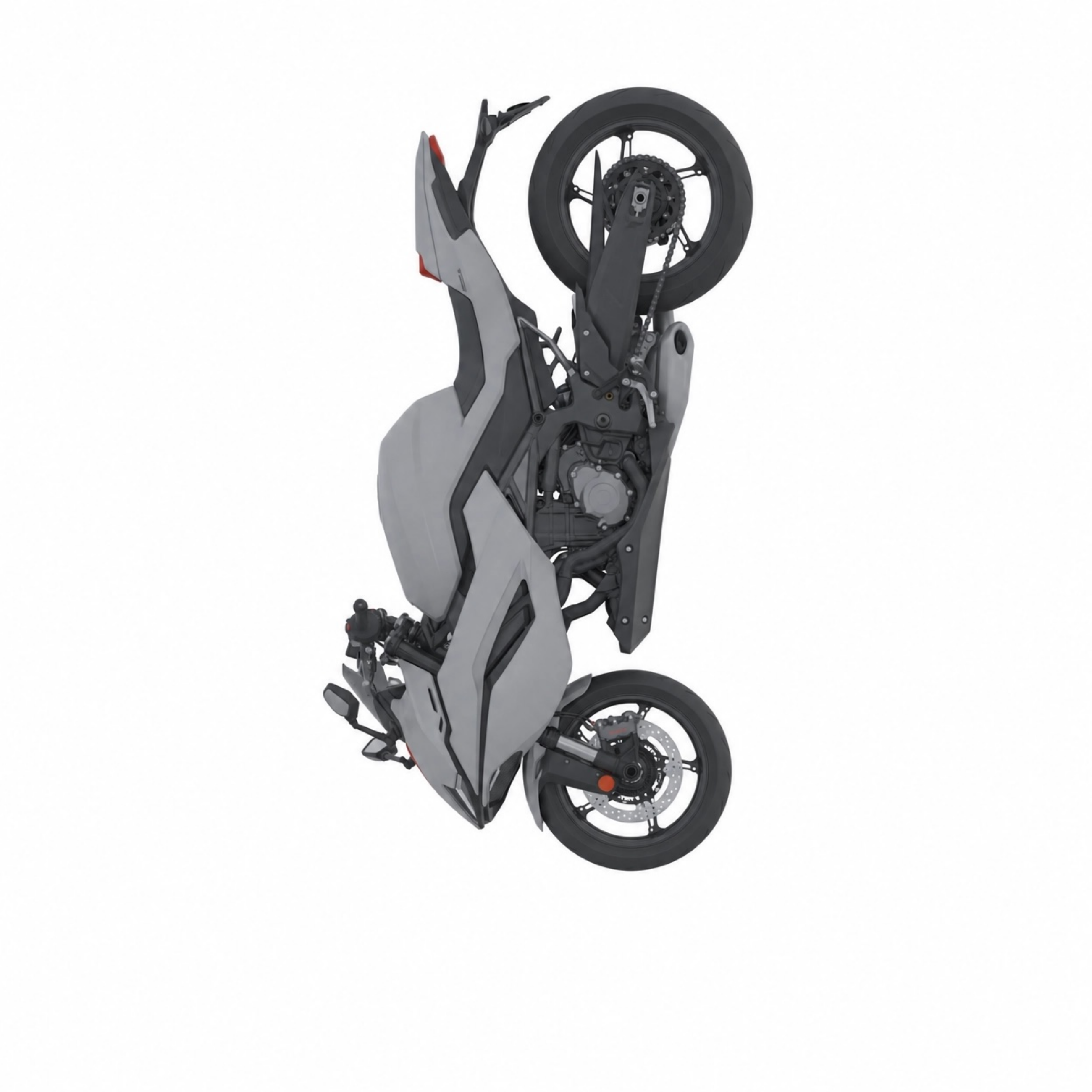} &
\includegraphics[width=0.24\linewidth]{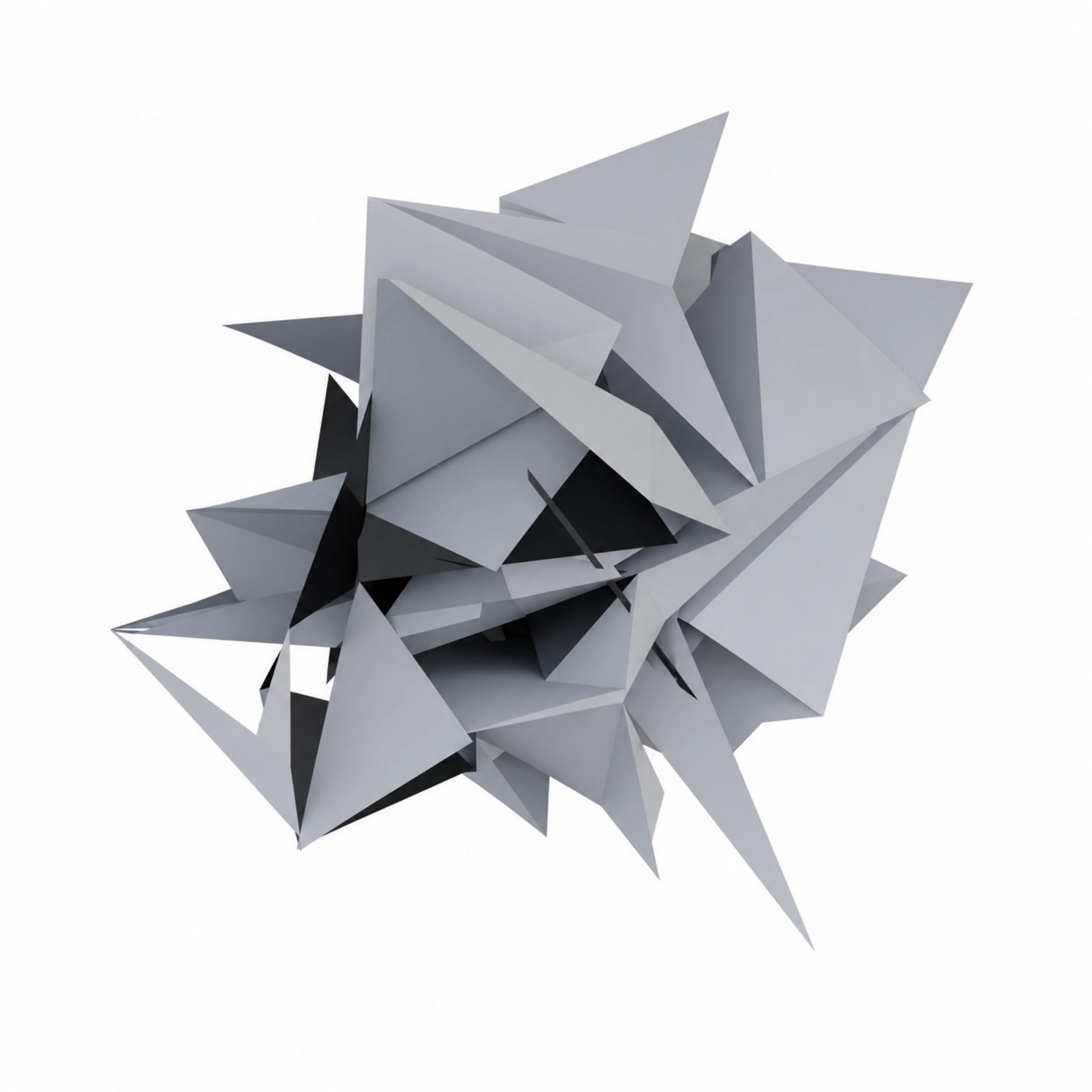} \\
{\small\textit{Background}} & {\small\textit{3D scan}} & {\small\textit{Wrong orientation}} & {\small\textit{Low quality}}\\
\end{tabular}

\caption{Examples of assets removed during semantic filtering: background geometry, 3D scans, incorrect orientation, and low mesh quality.}%
\Description{Data filtering examples}%
\label{fig:data_filtering_examples}
\end{figure}

The quality of a 3D generator depends strongly on the quality of the training data, which we enforce via rigorous curation.
We started from a large collection of in-house licensed datasets, but identified issues such as duplicates, background presence, 3D scans with degenerate geometry, inconsistent orientation and low mesh quality, as shown in Fig.~\ref{fig:data_filtering_examples}.
We filtered out problematic assets using geometric rules and a Vision-Language Model (VLM).

\paragraph{Geometric Filtering}%
\label{sec:data_geometric}

We started by identifying and removing \emph{duplicates} (e.g., objects that only had minor texture or geometric differences), using metadata and vertex count to identify them and retaining only the variant with the largest file size.
We also removed meshes with very low or very large polygon counts, often corresponding to placeholders or 3D scans.
This filtered around $30\%$ of the assets.
Approximately 3\% of the assets contained \emph{animations}.
From these, we selected the rest pose and some additional representative poses using max-distance sampling to encourage diversity.

\paragraph{Semantic Filtering}%
\label{sec:data_semantic}

The remaining assets underwent a multi-stage, VLM-driven semantic filtering process.
Each stage targets a specific failure mode, removing roughly $30\%$ assets in total.
First, artist-created models often include \emph{background} geometry (e.g., skydomes, ground planes, and surrounding cylinders) that interferes with object-centric generation.
We employ a two-stage detection mechanism: a lightweight heuristic detector first flags candidates based on geometric properties (e.g., outlier size, detached planar components), followed by a VLM classifier to filter false positives.
Second, photogrammetric \emph{3D scans} and \emph{satellite imagery}, which typically exhibit jagged boundaries, noisy surfaces, and non-manifold topology, are identified and removed via a dedicated VLM pass, filtering roughly $4\%$ assets.
Third, a large fraction of meshes exhibit incorrect \emph{3D orientation}, appearing upside-down or sideways.
We render four canonical views at a fixed elevation and apply an orientation predictor using majority voting to determine the 3D orientation.
A VLM then verifies predictions falling outside a plausible elevation range.
Objects with confirmed incorrect gravity are first removed (approximately 2\%).
Then, all retained meshes whose predicted frontal view deviates from the canonical forward direction are re-aligned, ensuring global orientation consistency across the training set.
Finally, the VLM was used to score each asset based on \emph{geometric fidelity} and \emph{texture quality} (both on a 0--10 scale) and to flag visual artifacts such as rendering glitches, anomalous viewing angles, transparency errors, or residual backdrops.
Assets with severe geometric degradation (score $\leq 2$) or flagged artifacts were excluded.
This step removed approximately 20\% assets.

\section{MeshGen}%
\label{sec:meshgen}

Given a single input image, MeshGen reconstructs the object 3D shape in the form of a mesh.
This section details the training data, model architecture, and distillation strategy to accelerate inference.

\subsection{MeshGen Data Preparation}%
\label{sec:meshgen-preprocessing}

MeshGen relies on signed distance functions (SDFs), which require the identification of interior and exterior volumes to be well-defined. To do so, we first convert raw 3D assets into watertight meshes.

We found standard methods employing ray tracing~\citep{zhang2024clay} or winding numbers~\citep{zhao2025hunyuan3d} to be unreliable; ray tracing is slow and fails near holes and slits, while winding numbers degrade with inconsistent face orientations.
Therefore, we use an adaptation of the flood-fill algorithm~\citep{smith79tint} that is robust to these defects. 
We identify the exterior with voxel-grid flood filling, accelerated by a GPU Bounding Volume Hierarchy (BVH).
First, we mark a narrow surface band: voxels whose unsigned distance to the input surface is below a fixed threshold.
We then flood fill from the grid boundary, stopping at this band.
Small gaps in the band can cause leakage into the interior.
Dilation seals these gaps but shifts the surface boundary, which cannot be compensated by erosion without removing high-frequency boundary detail.

\begin{figure}[t]
\vspace{-2mm}
\centering
\includegraphics[width=\columnwidth]{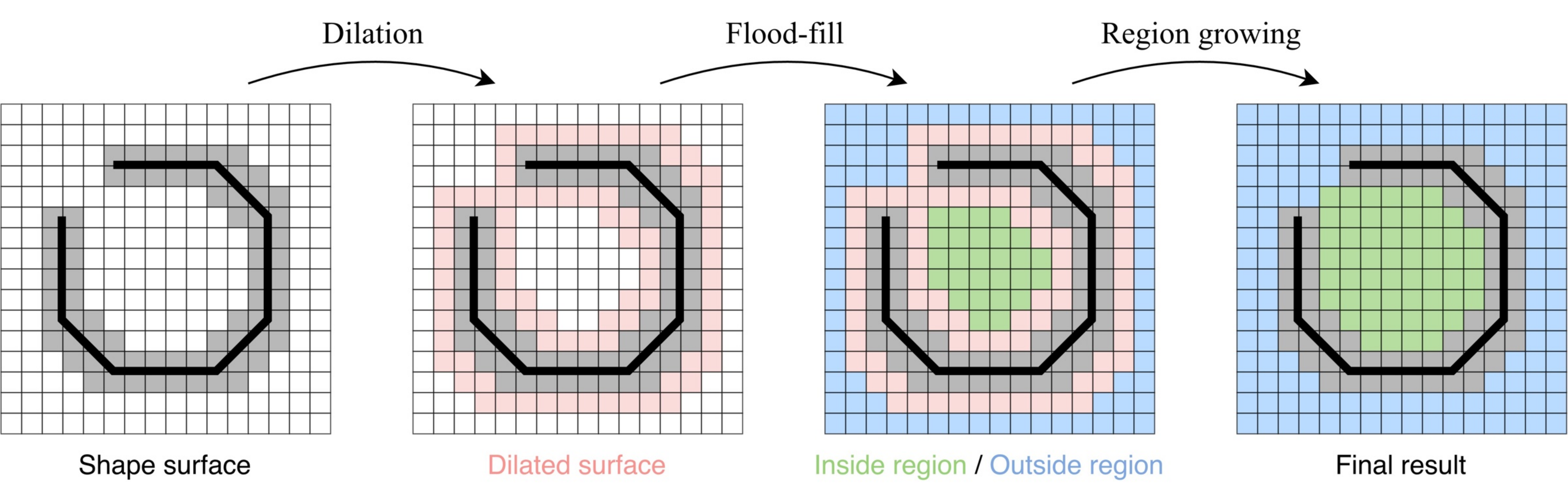}
\vspace{-6mm}
\caption{Overview of the flood-fill-based voxel sign estimation algorithm (shown in 2D cross-section).
See the text for details.
}%
\Description{Flood fill algorithm.}%
\label{fig:floodfill}
\end{figure}

We address this issue with a three-stage algorithm, illustrated in \cref{fig:floodfill}.
First, we dilate the surface band (gray) to close the gaps in the input mesh, producing an expanded band (pink).
Second, we flood fill from the voxel-grid boundary: flooded voxels are exterior (blue), dry voxels are interior (green), and voxels in the dilated band remain undetermined (pink).
Third, we resolve undetermined voxels by assigning the label that best agrees with their neighbors.
This replaces naïve erosion, while preserving high-frequency boundary detail. Since dilation can create small artificial cavities in concave regions or between nearby parts, we identify them as small connected components from the inferred interior and return them to the unknown set before final relabeling.

After labeling, we extract a watertight mesh with marching cubes and compute SDF supervision from the resulting surface for training.

\subsection{MeshGen Architecture}%
\label{sec:dit}

MeshGen generates 3D shapes from a single image using a latent diffusion transformer operating on the VecSet~\citep{zhang20233dshape2vecset} representation.
The model employs a coarse-to-refine architecture: an initial stage establishes the rough shape and topology of the object, and a refinement stage adds higher-resolution detail.

\paragraph{Formulation}

MeshGen follows a standard conditional latent diffusion formulation over VecSet shape latents. We summarize the formulation briefly, as the main system contribution lies in converting dense 3D generation into a latency-constrained, exportable asset pipeline.
Given an image $\img$, we generate a 3D shape with a diffusion transformer (DiT)~\citep{peebles2023scalable} operating in the VecSet latent space.
The VecSet Variational Auto-Encoder (VAE) encodes the input shape (initially represented as a set of 3D points with normals) into a compact set of $N$ latent tokens $\z \in \mathbb{R}^{N \times D}$; the decoder maps $\z$ to a corresponding SDF function, which can be queried at any spatial location to recover the shape.
For generation, we train a denoising model $\phi_\theta$ to reverse a forward diffusion process that progressively corrupts a clean latent code $\z_0 \sim q(\z)$ into Gaussian noise.
We adopt the \emph{v-prediction}~\citep{salimans2022progressive} parameterization with a scaled-linear beta schedule and zero-SNR rescaling~\citep{lin2024common}.
At each timestep $t$, the noisy latent $\z_t = \alpha_t \z_0 + \sigma_t \beps$ is constructed from the noise schedule coefficients $\alpha_t, \sigma_t$, and the model predicts the velocity $\mathbf{v}_t = \alpha_t \beps - \sigma_t \z_0$.
The training objective is
$
\mathcal{L}_\text{DM} = \mathbb{E}_{t, \beps, \z_0} \left[\left \| \phi_\theta(\z_t, t, \mathbf{c}) - \mathbf{v}_t \right \|^2\right]
$
where $\mathbf{c}$ denotes the image conditioning.

Classifier-free guidance (CFG)~\citep{ho2022classifier} is used to improve sample quality.
During training, the conditioning $\mathbf{c}$ is randomly dropped with probability $p_\text{cfg}$.
At inference, the guided prediction is computed as
$
\tilde{\mathbf{v}}_t = \mathbf{v}_t^\text{uncond}
+ s \cdot (\mathbf{v}_t^\text{cond} - \mathbf{v}_t^\text{uncond})
$
where $s$ is the guidance scale.
This requires two forward passes per denoising step (conditional and unconditional), which is subsequently eliminated via progressive distillation (\cref{sec:meshgen-distillation}).

\paragraph{Architecture}%
\label{sec:meshgen-architecture}

The backbone follows the DiT design~\citep{peebles2023scalable} with skip connections and AdaLN modulation.
It maps the noisy latent code $\z_t$ to the predicted velocity $\mathbf{v}_t = \phi_\theta(\z_t, t, \mathbf{c})$ through an input projection, a stack of DiT blocks, and an output projection.
A linear layer projects the $N \times D$ noisy latent tokens to the transformer hidden dimension $\dimt$, where $N$ denotes the number of latents.
The input image $\img$ is encoded by a frozen DINOv2 ViT-G/14 with registers~\citep{oquab2023dinov2,darcet2023vision}, producing a sequence of patch tokens $\mathbf{c} \in \mathbb{R}^{L \times D_c}$ at resolution 518px.
The transformer uses $n$ blocks, each consisting of three sub-layers with residual connections: (i) self-attention among latent tokens with AdaLN and query-key normalization, (ii) cross-attention to the image condition tokens $\mathbf{c}$ with AdaLN, and (iii) a SwiGLU feed-forward network~\citep{shazeer2020glu} with gated activation.
The final token representations are layer-normalized and linearly projected back to $N \times D$, predicting the velocity $\mathbf{v}_t$.

Our DiT is a 2.3B parameter model, it uses $n = 24$ blocks with hidden dimension $\dimt = 2048$, 16 attention heads and skip connecting the first/last 11 blocks.
The latent code at this coarse stage uses $N = 4096$ tokens of dimension $D = 64$.

\paragraph{Implementation details}

For training stability and efficiency, we train our model in three stages by gradually increasing the number of latent tokens $N$ from 512 to 4096.
In the first stage, we use $N = 512$ tokens and train for 750K iterations with a learning rate of $1 \times 10^{-4}$ and a batch size of 2048 samples.
In the second stage, we use $N = 2048$ and train for 500K iterations with a learning rate of $2 \times 10^{-5}$ and a batch size of 1280.
In the third stage, we use $N = 4096$ and train for 250K iterations with a learning rate of $1 \times 10^{-5}$ and a batch size of 1024.
For all stages, similar to the VAE training, we use AdamW~\citep{loshchilov2017decoupled}, a weight decay of 0.01, $\beta = (0.9, 0.95)$, and bfloat16 mixed precision.

\paragraph{Coarse-to-refine generation}%
\label{sec:twostage}

\begin{figure*}[t]
\centering
\includegraphics[width=1.0\textwidth]{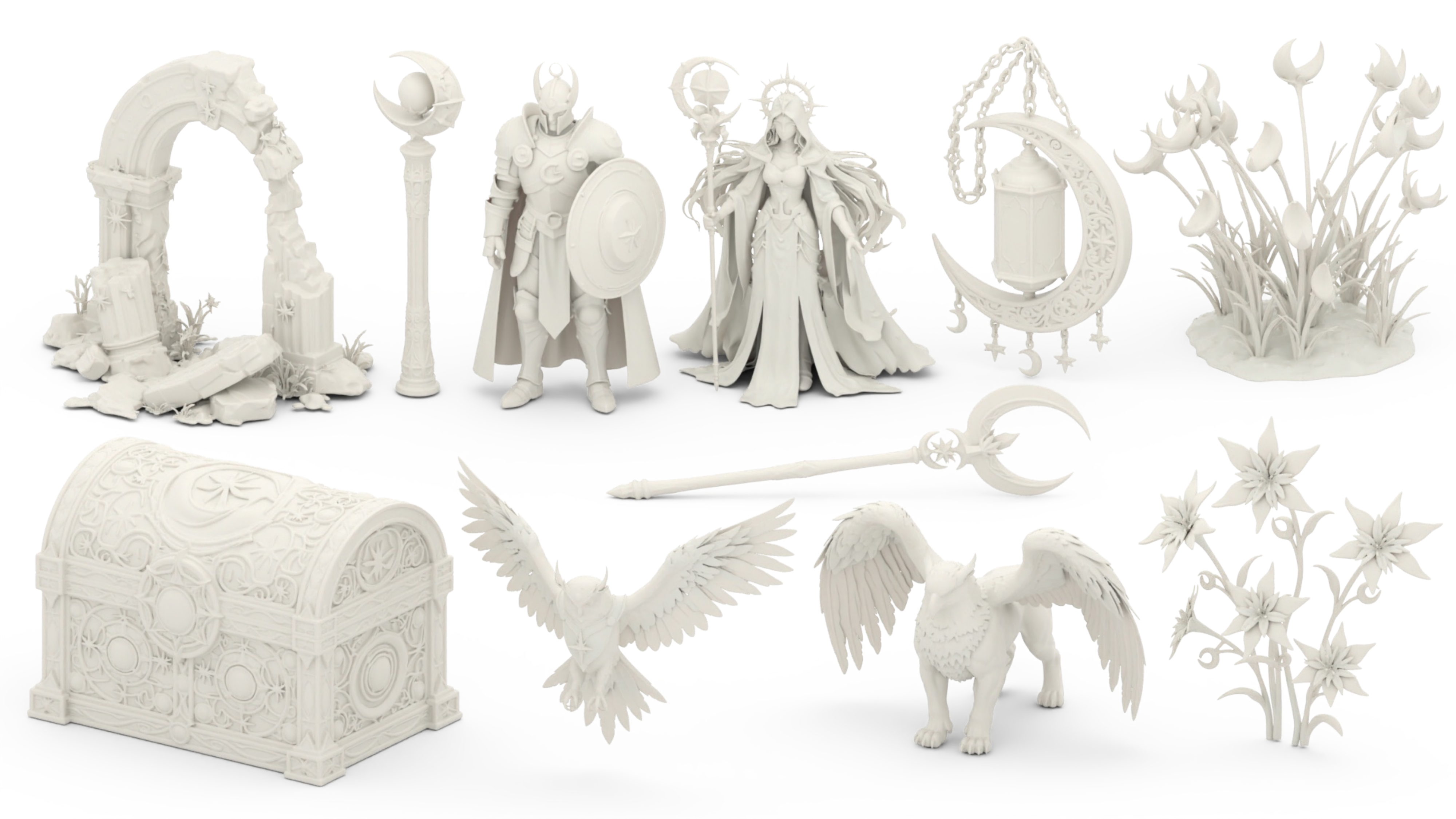}
\caption{Visualization of the dense meshes predicted by our two-stage coarse-to-refine MeshGen, prior to geometry processing and simplification.}
\label{fig:raw_mesh}
\end{figure*}

A single-stage model with $N = 4096$ latent tokens captures the overall shape but often lacks the resolution needed for fine geometric detail.
We recover them by introducing a second stage 
inspired by LATTICE~\citep{lai2025lattice}.

The refinement model takes the coarse mesh as input and predicts a higher-resolution latent code with $N_\text{refine} = 16{,}384$ tokens.
To construct the refinement tokens, we sample surface points from the coarse mesh and voxelize them to produce $N_\text{refine}$ 3D positions $\mathbf{x}_i \in \mathbb{R}^3$ for each latent token.
These positions serve as RoPE input in the refinement transformer's self-attention layers, anchoring each latent token to a specific location on the coarse geometry.
This is different from the coarse stage, where tokens have no preferential spatial association \emph{a priori}, and allows the model to focus more easily on local details rather than the global structure.

The refinement stage also uses higher-resolution image conditioning: the DINOv2 encoder operates at 1022px resolution, producing $4\times$ more patch tokens than in the coarse stage.
The predicted refined latent code is decoded through a dedicated refinement VAE trained with point jittering to make it robust to voxelization.
The decoder shares the coarse VAE interface---querying an implicit field and extracting the final surface via marching cubes---but uses the denser, spatially anchored token set to recover local surface detail around the coarse prediction.
Because the refinement stage is guided by the coarse geometry rather than solving global structure from scratch, it converges in fewer diffusion steps.

\subsection{MeshGen Distillation}%
\label{sec:meshgen-distillation}

At inference, the default MeshGen configuration uses 100 denoising steps for the coarse stage and 20 steps for the refinement stage, both with CFG, which further doubles the number of forward passes of the DiT.
We distill both models to accelerate inference while preserving quality.

We apply progressive distillation~\citep{salimans2022progressive} to each stage, reducing iterations from 100 to 25 and from 20 to 5 respectively (a $4\times$ reduction in denoising steps).
We also fold CFG into the distilled model to reduce compute and maximize speed: during training, the teacher uses guided sampling, and the student learns to reproduce the guided output in a single conditional forward pass.
To train the student, we use a DDIM-based target-inversion objective.
At each training step, the teacher starts from a randomly sampled student timestep and runs $N$ deterministic DDIM steps to produce a multi-step target.
The student is trained with an MSE loss to match this multi-step teacher output in a single step, using an inverted $v$-prediction target.
Optionally, we add an auxiliary $x_0$ reconstruction loss that penalizes deviation between the student's predicted clean latent and the ground truth, providing light regularization without changing the primary distillation objective.

Overall, on a single NVIDIA A100 80GB GPU, distillation reduces latency from \emph{13.9} to \emph{1.9} seconds for the coarse stage, and from \emph{14.3} to \emph{1.9} seconds for the refinement stage. To validate the distillation quality, we use human evaluation as well as quantitative metrics such as Chamfer distance do not sufficiently capture perceptual fidelity.
We curate 200 teacher-student output pairs spanning diverse categories (objects, characters, and environmental elements) and representative image-to-3D benchmark inputs, and present them in a randomized, blind side-by-side comparison. As summarized in \cref{tab:distillation_human_eval}, evaluators prefer the teacher in $32.9\%$ of comparisons, the student in $30.5\%$, and judge $36.6\%$ as similar, suggesting that a $4\times$ reduction in sampling steps incurs negligible perceptible quality loss.
See \cref{fig:raw_mesh} for example MeshGen meshes.

\begin{table}[ht]
\centering
\caption{Human evaluation of quality across 200 pairs (1018 total votes).
The $4\times$-distilled student closely matches the teacher in perceptual quality. Latency is measured on A100 GPU.}%
\label{tab:distillation_human_eval}
\setlength{\tabcolsep}{4pt}
\begin{tabular}{lccccc}
\toprule
Model & Steps & Sec./sample & Wins & Similar & Win Rate \\
\midrule
Teacher & 120 & 28.20 & 335 & 373 & 32.9\% \\
Student & 30 & 3.78 & 310 & 373 & 30.5\% \\
\bottomrule
\end{tabular}
\end{table}

\begin{table*}[ht]
\centering
\caption{Mesh simplification performance on 200 benchmark assets at 5K and 10K face targets. We evaluate geometric deviation (Chamfer, Hausdorff), orientation consistency (Flipped Normals across 24 views), and the mean angular error of both baked and unbaked geometric normals (Baked Mean, Geo Mean). All metrics are averaged over the benchmark ($\downarrow$ lower is better).}
\label{tab:simplification_comparison}
\scalebox{0.75}{
\setlength{\tabcolsep}{4pt}
\begin{tabular}{llcccccc}
\toprule
Target & Method & Latency (s) & Chamfer ($\times 10^{-5}$) $\downarrow$ & Hausdorff ($\times 10^{-3}$) $\downarrow$ & Flipped Normals $\downarrow$ & Baked Mean ($^\circ$) $\downarrow$ & Geo Mean ($^\circ$)
$\downarrow$ \\
\midrule
\multirow{3}{*}{5k}
& Houdini           & 14.64 & 36.5          & 5.19          & 3956          & \textbf{9.02} & 12.39          \\
& CuMesh (original) & 0.437 & 29.6          & 6.57          & 1593          & 10.58         & 14.94          \\
& CuMesh (ours)     & 0.469 & \textbf{13.3} & \textbf{4.59} & \textbf{1126} & 9.39          & \textbf{11.94} \\
\midrule
\multirow{3}{*}{10k}
& Houdini           & 11.58 & 29.7          & 3.87          & 997           & 6.60          & 9.39           \\
& CuMesh (original) & 0.429 & 8.70          & 1.90          & 454           & 7.02          & 10.78          \\
& CuMesh (ours)     & 0.457 & \textbf{3.01} & \textbf{0.97} & \textbf{268}  & \textbf{6.47} & \textbf{8.66}  \\
\bottomrule
\end{tabular}
}
\end{table*}

\section{Geometry Post-Processing for Runtime Assets}%
\label{sec:geometry_processing}
The output of MeshGen is an SDF, from which we extract an explicit surface mesh. 
We first run marching cubes at a high resolution ($512^3$) to obtain a dense mesh with roughly one million faces. 
This dense extraction preserves the high-frequency geometry and serves as the source geometry for subsequent processing. 
However, it is not directly suitable as a deployable asset: it is too dense for real-time rendering, may contain hidden interior surfaces, and does not provide the UV parameterization required for texture mapping.
We therefore convert it into a texture-ready asset through mesh simplification, visibility pruning, UV unwrapping, and tangent-space normal baking. The normal map transfers high-frequency surface detail from the dense source mesh onto the simplified mesh, yielding a compact representation suitable for real-time rendering on mobile.

\subsection{Mesh Simplification}%
\label{sec:simplification}

We considered using production-grade mesh simplification software, but found existing solutions too slow for our purposes.
We therefore develop our own solution.

We build on CuMesh\footnote{\url{https://github.com/visualbruno/CuMesh}}, a library that implements parallel edge collapse with quadric error metrics (QEM)~\citep{garland1997surface,oh2025pamo} that runs in less than one second on a GPU\@ for our target mesh size.
The algorithm maintains a quadric $Q_v$ for each vertex $v$, encoding the squared distance
$
E(\mathbf{v}) = [\mathbf{v};1]^\top Q_v [\mathbf{v};1]
$
of a point $\mathbf{v}$ to all incident face planes.
The cost
$
\min_{\mathbf{v}}  E(\mathbf{v})
$
of collapsing an edge $(v_0, v_1)$ is computed from the combination $Q = Q_{v_0} + Q_{v_1}$ of the quadrics of its two endpoints.

The baseline CuMesh implementation does not solve for the QEM-optimal position;
it uses $Q$ only to rank candidate collapses, then places the contracted vertex at the midpoint (interior edges) or an endpoint (boundary edges).
This means the collapse ordering reflects the QEM objective, but the actual vertex positions do not, leading to systematic geometric degradation under aggressive polygon reduction.
We compute the contraction position from the merged quadric directly, with an edge-constrained projection for stability.
For interior edges, we compute the unconstrained QEM minimizer $\mathbf{v}_\text{opt}$ of the energy $E(\mathbf{v})$.%
\footnote{
I.e., we solve $A\mathbf{x} = -\mathbf{b}$ (where $A$ is the upper-left $3{\times}3$ block of $Q$ and $\mathbf{b}$ its upper-right $3{\times}1$ column).
}
We then project $\mathbf{v}_\text{opt}$ onto the edge segment $\mathbf{v}(t) = (1{-}t)\,\mathbf{v}_0 + t\,\mathbf{v}_1$ by computing
$
t = \langle \mathbf{v}_\text{opt} - \mathbf{v}_0,\, \mathbf{v}_1 - \mathbf{v}_0 \rangle / \lVert \mathbf{v}_1 - \mathbf{v}_0 \rVert^2
$
and clamping $t \in [0,1]$.
This bounds vertex displacement to the edge length while following the QEM objective.
When the quadric is singular, we fall back to the midpoint.
For boundary edges, we keep boundary vertices fixed or select the lower-error endpoint.
The same placement logic is used for both cost evaluation and actual collapse, ensuring the ordering reflects the true geometric cost.

\paragraph{Mesh simplification results}

We evaluate the impact of optimal placement on a benchmark of 200 meshes generated by our image-to-3D pipeline, comparing three configurations: Houdini's PolyReduce as a representative production-quality CPU solver, CuMesh with the baseline midpoint placement (original), and CuMesh with our optimal placement strategy (ours).
Each mesh is simplified to two target face counts (5k and 10k) and compared against the raw mesh.

We measure the following metrics:
The \emph{Chamfer distance} samples 256k surface points from the original and simplified meshes and reports the average minimum distance of points in one set against the other, in both directions.
The \emph{Hausdorff distance} does the same but reports the maximum minimum distance.
The \emph{Flipped normal} metric renders the simplified mesh from 24 viewpoints with and without backface culling at resolution of 1024 and reports the number of pixels where the two renderings disagree, caused by a flipped normal.
\emph{Baked mean} bakes the normals from the original mesh to the simplified one and reports the average angular error in degrees of the approximation.%
\footnote{The simplified mesh is UV-unwrapped via xatlas, and a normal map is baked from the raw mesh onto the simplified mesh's UV layout.
Surface points are then sampled uniformly on the simplified mesh, and the angular error between the baked normal and the true raw-mesh normal is computed at each sample.
The metric reports the mean error in degrees across all samples.}
\emph{Geo mean} reports the average angular error between the simplified and original mesh true normals.

Results are summarized in \cref{tab:simplification_comparison}. 
The optimized placement improves over the CuMesh baseline across all metrics.
Compared with Houdini, the optimized CuMesh achieves better geometric fidelity on most measures---Chamfer distance, Hausdorff distance, flipped normals, and geometric-normal error---with Houdini slightly ahead only on baked-normal error at the 5k-face target.

\subsection{Hidden Face Removal}%
\label{sec:hidden_face_removal}

Generated meshes can contain interior faces that are invisible from any exterior viewpoint. Such faces waste polygon and texture budget, complicate UV unwrapping, and may create downstream artifacts in rendering or rigging. 
We remove them with a GPU-accelerated visibility pass followed by topology-aware cleanup.

The visibility pass builds a BVH over the mesh and casts rays from 512 viewpoints distributed on a surrounding sphere at 1024-pixel resolution. A face is marked visible if at least one ray reaches it without occlusion. The BVH is constructed once and reused across all viewpoints, and the result is accumulated directly into a per-face visibility mask.

Pure visibility classification produces fragile results under self-occlusion: rays can miss faces that belong to the exterior surface but are shadowed by other geometry, leaving isolated floating triangles or holes in the visible shell.
A topology-aware flood-fill step addresses this by propagating exterior classification across the mesh's connectivity.
The flood fill is seeded from faces already classified as visible by ray tracing.
Propagation proceeds via breadth-first traversal across edges, gated by a normal consistency check: an adjacent face is promoted to exterior only if its normal agrees in direction with the seed face, preserving winding consistency and preventing propagation through the interior.

\paragraph{Results}
We evaluate the pipeline on 100 meshes produced by our generation model, each containing approximately 22K faces after simplification. The core hidden-face-removal kernel runs in 0.179 seconds per mesh on average on a single GPU. In the production pipeline, the corresponding stage averages 0.362 seconds once staging and scheduling overhead are included.

\subsection{UV Unwrapping}%
\label{sec:uv_unwrapping}

After simplification and hidden face removal, we construct a UV atlas for the mesh to support normal baking and texturing in subsequent stages.
We use xatlas~\footnote{\url{https://github.com/jpcy/xatlas}} for UV unwrapping.
It operates in three stages: chart segmentation, which partitions the mesh surface into regions suitable for planar mapping; LSCM-based parameterization, which computes low-distortion 2D coordinates for each chart; and atlas packing, which arranges charts into a texture atlas.

Of these, chart segmentation dominates runtime because it grows charts one face at a time with a greedy priority queue over normal-compatibility costs, serially.
We first evaluated CuMesh's GPU-based chart segmentation\footnote{\url{https://github.com/visualbruno/CuMesh}}, which uses parallel edge collapse with normal-cone merging.
However, this rebuilds the graph after each round of edge collapse, causing frequent synchronization across small GPU kernels and leads to no acceleration for meshes with fewer than 30k faces as the kernel launches become the bottleneck.

We instead introduce a parallel chart segmentation pipeline that partitions the mesh spatially, runs segmentation on each partition with boundary overlap, and reconciles the results through a merge pass.
The approach follows a four-phase partition-with-overlap strategy.
First, we recursively bisect the mesh into spatially balanced regions using median-area splits along the axis of maximum centroid spread, producing a set of compact macro-partitions.
Second, we extract an N-ring halo of neighboring faces around each partition boundary via breadth-first traversal, so that each partition's segmentation has geometric context about adjacent regions and does not produce arbitrary chart boundaries at partition edges.
Third, we run the full xatlas segmentation pipeline independently on each partition (core faces plus halo) in parallel using the existing task scheduler.
Fourth, we reconcile the per-partition results into a unified chart assignment through a three-step boundary reconciliation: core faces are assigned to charts from their owning partition, orphan faces (those in halo zones that were not claimed) are assigned to the neighboring chart with the best normal compatibility, and a final pairwise merge pass---ported from xatlas's serial merge logic with the same five merge criteria (single-face absorption, quad absorption, enclosed chart, significant shared boundary, and normal gate)---consolidates fragmented charts at partition boundaries while validating that merged normals remain coherent.

\paragraph{Human Evaluation}

We conducted a human evaluation of final texture quality, where the only variable between conditions is the UV unwrapping algorithm---all other pipeline stages (mesh generation, simplification, and texture synthesis) are held constant.
Across two evaluation rounds totaling 604 pairwise comparisons, the combined results are summarized in \cref{tab:xatlas_human_eval}.
The serial baseline holds a marginal advantage at 51.9\% win rate, while the parallel unwrap provides a 1.9$\times$ speedup on average.
Annotators reported that distinguishing the two methods required zooming in to inspect seam boundaries, and that the differences were not visible on mobile devices.
This makes UV parameterization a good Pareto trade: a visible latency reduction with no practically observable degradation for interactive use.

\begin{table}[htbp]
\centering
\caption{Human evaluation of UV unwrapping quality comparing serial and parallel \textit{xatlas} across 604 pairwise comparisons. Processing speed is normalized to the serial baseline.}%
\label{tab:xatlas_human_eval}
\setlength{\tabcolsep}{4.3pt}
\small
\begin{tabular}{lccccccc}
\toprule
Xatlas & Speed (s) & Wins & Losses & Ties & Total & Win \\
\midrule
Serial & 4.70 & 241 & 217 & 146 & 604 & 51.9\% \\
Parallel & 2.56 & 217 & 241 & 146 & 604 & 48.1\% \\
\bottomrule
\end{tabular}
\end{table}

\subsection{Normal Map Baking}%
\label{sec:normal-baking}

The simplification stage retains the raw high-poly mesh alongside the simplified output.
We bake a tangent-space normal map that transfers high-poly surface detail onto the simplified mesh's UV layout, recovering geometric fidelity that would otherwise be lost to polygon reduction.

Blender’s Cycles renderer produces high-quality normal maps but averages 2.92s per mesh in our benchmark and can take up to 25.85s in the worst case.
It also incurs data-conversion overhead: the mesh must be exported for Blender to load, and the baked output must be saved and read back, adding further latency.
We replace Blender with a GPU-resident pipeline that keeps all data in memory, eliminating inter-process overhead.
The pipeline has four stages.

The pipeline rasterizes the low-poly mesh into its own UV space using DRTK~\citep{pidhorskyi2024rasterized}\footnote{\url{https://github.com/facebookresearch/DRTK}}, producing a G-buffer with world-space position, smooth normal, tangent, and bitangent for each valid texel.
A GPU BVH built on the high-poly mesh then locates, for each texel's world-space position, the closest point on the high-poly surface and retrieves its interpolated smooth normal via barycentric weighting.
Two filters reject erroneous correspondences: a distance filter discards hits beyond 1\% of the bounding-box diagonal, and a UV-reliability filter marks texels on faces with degenerate UV parameterization as unreliable, assigning them a neutral tangent-space normal $(0, 0, 1)$.
The hit normal is projected into the texel's tangent frame, encoded as RGB8, and dilated at UV seam boundaries to prevent mipmap bleed.
The entire pipeline runs on the GPU with no disk I/O.

\begin{table}[ht]
\centering
\caption{Comparison of normal baking performance between our GPU-resident implementation and Blender. Evaluated across 50 benchmark assets at a target of 5K faces and $1024 \times 1024$ texture resolution.}
\label{tab:normal-bake-comparison}
\begin{tabular}{llccc}
\toprule
Metric & Method & Mean & Median & Max \\
\midrule
\multirow{2}{*}{Bake latency (ms)}
& Blender &   2{,}920 &   1{,}842 & 25{,}854 \\
& GPU     & \textbf{271} & \textbf{188} & \textbf{1{,}147} \\
\midrule
\multirow{2}{*}{Baked mean (\degree)}
& Blender &    9.89 & \textbf{6.98} &   54.39 \\
& GPU     & \textbf{9.66} &    7.14 & \textbf{50.35} \\
\midrule
\end{tabular}
\end{table}

\paragraph{Evaluation.}
We evaluate normal baking on 50 meshes, each simplified to 5k faces and paired with a raw high-poly source, as shown in~\Cref{tab:normal-bake-comparison}.
All methods use xatlas UVs and $1024 \times 1024$ tangent-space normal maps.
Our GPU baker achieves comparable quality to Blender while substantially reducing latency.
It runs $10.8\times$ faster on average (271\,ms vs.\ 2{,}920\,ms) and $22.5\times$ faster in the worst case (1{,}147\,ms vs.\ 25{,}854\,ms).
The mean angular error is slightly lower than Blender ($9.66^{\circ}$ vs.\ $9.89^{\circ}$), while Blender has a small median advantage ($6.98^{\circ}$ vs.\ $7.14^{\circ}$).

\section{TextureGen}
\label{sec:texturegen}%

The geometry stage outputs a simplified mesh, a UV atlas, and a tangent-space normal map that transfers high-frequency detail from the dense source.
The next step is TextureGen, which generates a UV texture for the object.

TextureGen decomposes the problem into two stages: image-space appearance synthesis and UV-space geometric fusion.
First, a multi-view diffusion transformer generates a coherent set of textured views conditioned on rendered geometry, the reference image, and text.
Second, a projection module fuses these views into the UV atlas and completes unobserved regions.
This separation leverages strong image-space diffusion priors while keeping the final texture geometrically aligned to the mesh, following recent work~\citep{bensadoun2024meta,cheng2025mvpaint}.
For interactive production use, we combine multimodal conditioning with efficient multi-view attention (\Cref{sec:tg_diffusion}), progressive distillation from a 32-step teacher to 4 steps (\Cref{sec:distillation}), and fast UV backprojection and completion (\Cref{sec:tg_backprojection}). See \cref{fig:texturegen_gallery} for example TextureGen results.

\subsection{TextureGen Data Preparation}

The TextureGen generator takes as input the 3D mesh and several target viewpoints, encoded as set of normal and position map renders, along with a reference image capturing the object appearance, and text prompt describing the object.
It generates as output several images of the textured object, one for each target view.
We explain below how the training data is obtained from each 3D asset.

\paragraph{Images.}

To obtain the \emph{reference image}, we render the 3D assets from a random camera viewpoint with azimuth and elevation in the range $-45^\circ$ to $45^\circ$ sampled in five discrete increments.
For each viewpoint, we render the mesh in two variants, one illuminated by a randomly selected and oriented HDRI image from a set of HDRI images with varying degrees of directional strengths, and one by a relatively diffuse HDRI image.
This makes the model robust to passing reference images that range from harsh directional lighting to flat diffuse appearance.

To obtain the \emph{target images}, we render the 3D asset from ten fixed viewpoints: 8 side views at $45^\circ$ azimuth increments, along with top and bottom views.
We use orthographic projection because it decouples projection scale from depth, yielding predictable foreground coverage for normalized assets. It also gives side views a shared vertical coordinate: pixels in the same image row correspond to the same 3D height, which simplifies learning cross-view consistency.

We scale the mesh to fit in a fixed sphere and render world-space normals and positions for each view.
We apply normal maps that come with the asset so the conditioning signal retains micro-surface details.
Position maps store the visible 3D coordinate at each pixel.

We task TextureGen with generating a color texture (instead of PBR materials) as this is more friendly with runtime rendering on mobile devices. 
We thus optimize the color targets to be visually appealing and informative, including soft shadows, ambient occlusion, and surface gradients, while avoiding view-dependent effects, such as specular reflections, that cannot be represented by a color texture.

Rendering the asset using a softly-lit environment is visually appealing, but results in view-dependent effects.
Those are removed by using diffuse illumination, but the latter results in a flat appearance that does not capture well the object shape and material.
We thus randomly blend these two renders, where the same weights are used across all 10 views and are sampled in the range from 0.2 (i.e., 20\% diffuse image) to 0.5 for each training step.

\paragraph{Text Prompts}%
\label{sec:tiled_captioning}

We use a Vision-Language Model (VLM) to caption the asset and obtain a corresponding text prompt.
To do so, we compose four renders into a $2\times2$ grid and feed it to the VLM, prompting it to describe the object as a whole.
Then, we flag poor descriptions that contain refusals, prompt echoes, meta-references, or insufficient content and ask an LLM to rewrite those preserving only factual object descriptions.
Finally, we task the LLM with summarising captions to a length of at most 128 tokens.

\subsection{TextureGen Architecture}%
\label{sec:tg_diffusion}

The core of TextureGen is a multi-view diffusion transformer that synthesizes a coherent set of color images from a reference image, text prompt, and per-view geometric renderings.
Let $\mathbf{I} \in \mathbb{R}^{H \times W \times 3}$ denote the reference image, $\mathcal{M}$ the processed mesh, and $\{\mathbf{N}_k, \mathbf{P}_k\}_{k=1}^K$ the normal and position maps rendered from $K$ views.
Given these inputs, the model generates textured views $\{\mathbf{V}_k\}_{k=1}^K$ simultaneously.
We describe its latent formulation, conditioning design, and structured multi-view attention.

\paragraph{Formulation.}

TextureGen operates in the latent space of a pretrained image VAE\@.
Each $1024 \times 1024$ view is encoded as a $64 \times 64$ latent.
The reference image and the $K$ target views are arranged into a $1 \times (1{+}K)$ latent grid: the first slot contains the reference image, and the rest are reserved for noisy target-view latents.
The latent in the reference slot is pinned throughout denoising, so the model treats it as a fixed appearance anchor.
The training loss is applied only to the $K$ target-view slots.

We train the model with optimal-transport conditional flow matching~\citep{tong2023improving,kornilov2024optimal} and a uniform timestep schedule.
The model predicts a velocity field $\mathbf{v}_\theta(\mathbf{z}_t, t, \mathbf{c})$, where $\mathbf{z}_t$ is the noisy multi-view latent at time $t$ and $\mathbf{c}$ collects the geometric renders, pinned reference image, and text conditioning.

\paragraph{Architecture.} The backbone is a diffusion transformer with 30 layers, hidden dimension 3072, 24 attention heads, and SwiGLU feed-forward layers with $4\times$ expansion.
Each block applies AdaLN-Zero modulation from the diffusion timestep, followed by self-attention, cross-attention to text embeddings, and a feed-forward layer.
The transformer processes the reference and target-view latents jointly, allowing appearance cues from the reference slot and geometric cues from each target view to interact during denoising.

For each target viewpoint, the rendered normal map $\mathbf{N}_k$ and position map $\mathbf{P}_k$ are encoded at the latent resolution and concatenated with the noisy target-view latent along the channel dimension.
This provides dense geometric context:
surface orientation from normals and surface correspondence from world-space positions.

The pinned reference slot provides the main appearance signal, including color, material, and style.
Because it is part of the same latent grid as the generated views, the model can propagate reference appearance through self-attention.
For CFG, the unconditional branch replaces the reference image with a uniform-white latent, allowing the guidance scale to control adherence to the visual reference.

Text prompts provide complementary semantic information that may be ambiguous from the reference image alone.
Text embeddings from T5 and CLIP are injected through cross-attention in each transformer block.

\paragraph{Structured multi-view attention.}%
\label{sec:tg_attention}

Full self-attention over $(1{+}K)$ latent slots is expensive when $K{=}10$ and each slot contains $64 \times 64$ tokens.
Instead, we use a sparse attention pattern matched to the camera layout.
Each target view attends to the reference slot and a small set of adjacent views.
Cardinal side views attend to the reference and their two neighboring views; diagonal views attend to the reference and neighboring cardinal views; top and bottom views attend to the reference and all four cardinal views.
Learned per-slot positional offsets identify each view within the grid.

For $K{=}10$, full attention would involve 45{,}056 tokens across the reference and target views.
The structured pattern reduces the per-view attention context by roughly $4\times$ while preserving the connections needed for cross-view consistency: every view sees the reference image, and neighboring views share overlapping surface regions.
\subsection{TextureGen Distillation}%
\label{sec:distillation}

Similarly to \Cref{sec:meshgen-distillation}, we use progressive distillation~\cite{salimans2022progressive} for TextureGen that progressively reduces denoising steps $32{\rightarrow}16{\rightarrow}8{\rightarrow}4$. We apply this to two model configurations: the default, \assetgenpro, uses a 4-step sampler (with CFG during sampling) to produce highest-fidelity textures at $1024{\times}1024$ resolution. \assetgenflash takes an extra stage to fold CFG into the distilled model to generate per-view at $768{\times}768$ resolution.

\paragraph{Progressive Distillation.}

The training objective combines velocity matching with an auxiliary denoised-image prediction loss.
After each stage, we select the best student checkpoint using pixel-level fidelity metrics, including PSNR and SSIM, against the immediate teacher on a benchmark suite.

\paragraph{CFG Distillation.} 

Following the same protocol as progressive distillation above, we select the best student checkpoint via PSNR against the 4-step CFG teacher.
We find that CFG distillation is particularly sensitive to the loss function.
The baseline MSE loss establishes 30.93~dB.
Switching to a Pseudo-Huber loss leads to a $+$1.08~dB improvement.
Adding channel-wise statistics matching and timestep importance sampling reduces color drift and stabilizes early training, though it does not improve PSNR.
Combining these with dynamic thresholding and a guidance-scale curriculum gives a further $+$0.13~dB, reaching 32.14~dB.

\subsection{TextureGen Postprocessing}%
\label{sec:tg_postprocess}%
\label{sec:tg_backprojection}

Given the $K$ texture views generated by TextureGen, the next step is to transfer them onto a single UV texture map via backprojection and blending followed by inpainting to fill unobserved regions.

\paragraph{Backprojection}%
\label{sec:backprojection_blending}

We first project the $K$ generated views to the texture UV atlas using \emph{backward} projection%
\footnote{Forward projection tends to leave holes and be inaccurate, particularly for grazing angles.}.
We consider each texel in the atlas, recover its corresponding 3D surface position and project that point into each of the $K$ generated  views, reading off the corresponding color.

Pixels near object boundaries or depth discontinuities are often unreliable, since small projection errors can sample background colors or colors from the wrong surface.
We therefore compute an edge mask by detecting large differences between neighboring pixels in the point coordinate maps, excluding them from backprojection.

A single bilinear sample per texel can undersample the generated view, especially at grazing angles and near UV seams where the UV-to-image mapping is highly anisotropic.
We address this issue by using per-texel anisotropic filtering.
For each atlas texel, we estimate its image-space footprint using finite-difference Jacobians of the UV-to-image mapping.
An SVD of this footprint gives the major and minor axes: the minor axis selects the mip level, while the major axis defines a multi-sampling direction with up to 8 taps.
The mip chain is built using a Lanczos-2 downsampler.
We apply a small negative LOD bias ($-0.5$), per-level sharpening, trilinear blending between adjacent mip levels, and anisotropy-aware mip compensation to retain detail along elongated footprints.

\paragraph{Blending.}
At this point, each texel has zero, one, or more candidate colors from the $K$ backprojected views.
For each texel with at least one candidate, we blend the contributions using an incidence-weighted scheme.

For each view $k$, we compute an incidence map $I_k \in [0,1]^{H_\text{uv} \times W_\text{uv}}$ in UV space.
Each texel is assigned a reliability score based on its surface orientation and visibility from the camera.
The raw incidence is the clamped dot product between the surface normal $\mathbf{n}$ and the viewing direction $\mathbf{d}_k$: $I_k^\text{raw} = \max(0, \mathbf{n} \cdot \mathbf{d}_k)$.
We then apply a depth-buffer visibility test.
The texel's 3D position is projected into view $k$, and its projected depth is compared against the rasterized depth buffer.
If the texel is occluded beyond a fixed tolerance, its incidence is set to zero.
Thus, $I_k$ is nonzero only for texels that are both front-facing and visible in view $k$.

Backprojecting view $k$ produces a partial atlas $A_k$.
We blend the partial atlases using the incidence maps and a per-view prior:
$T = \sum_{k=1}^{K} w_k \, I_k^{\,\alpha} \odot A_k \,\big/\, \bigl(\sum_{k=1}^{K} w_k \, I_k^{\,\alpha} + \epsilon\bigr)$,
where $\alpha$ controls the sharpness of the blend and $w_k$ encodes a view prior.
We assign higher weights to perceptually important views: $w_k=1$ for the front and rear views, $0.3$ for top and bottom, $0.1$ for the $\pm45^\circ$ frontal side views, $0.01$ for left and right side views, and $0.001$ for the $\pm45^\circ$ rear side views.
Together, the incidence term and view prior favor visible, near-orthogonal observations from salient viewpoints while suppressing grazing and less informative contributions.
We implement the reduction in log space using \texttt{logsumexp} for numerical stability when $\alpha$ is large.

\paragraph{Inpainting}

Texels that are not reliably observed from any generated view remain unfilled after blending.

Following MVPaint~\citep{cheng2025mvpaint}, we use the 3D geometry to guide texture completion in UV space.
Given the blended atlas, a binary missing-region mask, and per-texel position and normal maps, we build a KD-tree over known texels in 3D position space.
For each unknown texel, nearest-neighbor candidates are retrieved in 3D and weighted by spatial proximity and normal alignment.
Colors are then propagated from known to unknown texels using these geometry-aware affinities.
This step fills in occluded regions using texels that are nearby on the object surface, rather than merely nearby in UV space.

\subsection{Geometric Condition Rendering}%
\label{sec:condition-rendering}
One large remaining 
latency source for TextureGen is multi-view geometric condition rendering.
During training, we use Blender to render both geometric conditions and color targets from a shared scene setup.
However, Blender takes approximately 5\,s to render 10 views.
At inference, we replace Blender with DRTK~\citep{pidhorskyi2024rasterized}, a GPU-accelerated rasterizer that
renders all 10 views directly as GPU-resident tensors in approximately 500\,ms.
In a blind evaluation, these two rendering methods produce no visible difference in the final generated textures; DRTK-based results are preferred 54.3\% of the time over Blender-based results on 100 pairs.

\section{Latency Optimization}%
\label{sec:latency-optimization}

We have explained how distillation reduces the cost of sampling geometry and texture, but many other factors determine end-to-end latency: surface extraction, simplification, hidden-face removal, UV parameterization, normal baking, multi-view condition rendering, backprojection, inpainting, asset serialization, and I/O.
These stages stress different parts of the system: diffusion sampling saturates GPU tensor cores; geometry and texture preparation mix GPU kernels with CPU-side scheduling, KD-tree construction, and UV bookkeeping; and input download, output upload, image encoding, and asset serialization are I/O-bound.

We therefore optimize the system at two levels.

First, kernel- and precision-level optimizations reduce the cost of individual model inferences (\cref{sec:kernel-precision}).
Second, pipeline-level scheduling overlaps independent CPU, GPU, and I/O work so that request latency is closer to the critical path than to the sum of all stages (\cref{sec:pipeline-parallel}).
With these optimizations, \assetgenpro runs in approximately 30 seconds end to end, and \assetgenflash in 14 seconds end to end (\cref{tab:latency-breakdown-meshgen,tab:latency-breakdown-texturegen}).

\begin{table}[ht]
\centering
\caption{Latency breakdown of the MeshGen and geometry processing stages for AssetGen and \assetgenflash. Values are averaged over 100 benchmark images. A dash (--) indicates a stage skipped by the Flash configuration. Row values represent individual stage runtimes; \textit{the reported total reflects the measured critical-path latency rather than an arithmetic sum.}}
\label{tab:latency-breakdown-meshgen}
\small
\setlength{\tabcolsep}{3.5pt}
\begin{tabular}{lcc}
\toprule
\textbf{Module} 
& \cellcolor{assetgenfullrow}\textbf{AssetGen (s)} 
& \cellcolor{assetgennanorow}\textbf{Flash (s)} \\
\midrule
Preprocess & 0.227 & 0.227 \\
Coarse conditioning & 0.044 & 0.044 \\
Coarse diffusion & 1.193 & 1.193 \\
Coarse mesh decoding & 1.414 & 1.414 \\
Coarse surface sampling & 0.636 & --- \\
Refine conditioning & 0.054 & --- \\
Refine diffusion & 1.671 & --- \\
Refine mesh decoding & 1.704 & --- \\
Floater removal + simplification & 0.654 & 0.654 \\
Hidden-face removal & 0.362 & 0.362 \\
UV unwrapping & 2.561 & 2.561 \\
Tangent normal baking & 0.265 & 0.265 \\
\midrule
Total MeshGen + geometry critical path
& \cellcolor{assetgenfullrow}\textbf{10.829} 
& \cellcolor{assetgennanorow}\textbf{6.768} \\
\bottomrule
\end{tabular}
\end{table}

\begin{table}[ht]
\centering
\caption{Latency breakdown of the TextureGen stage. Values are averaged over 100 benchmark images. The default AssetGen configuration uses 1024px diffusion with inference-time CFG (split across two GPUs), per-view super-resolution, and a 2K texture atlas. \assetgenflash uses 768px diffusion with a guidance-distilled conditional-only model, omits super-resolution, and exports a 1K atlas. Row values represent individual stage runtimes, while \textit{the reported total reflects the measured critical-path latency rather than an arithmetic sum.}}%
\label{tab:latency-breakdown-texturegen}
\setlength{\tabcolsep}{3.5pt}
\small
\begin{tabular}{lcc}
\toprule
\textbf{Module} 
& \cellcolor{assetgenfullrow}\textbf{\assetgenpro (s)} 
& \cellcolor{assetgennanorow}\textbf{\assetgenflash (s)} \\
\midrule
Preprocess & 0.346 & 0.228 \\
Render conditions & 0.642 & 0.394 \\
Diffusion & 10.455 & 4.420 \\ 
Precompute backproject data & 1.455 & 0.570 \\
Build KD tree & 0.908 & 0.274 \\
Per-view super-resolution & 3.112 & --- \\
Backprojection and inpainting & 0.801 & 0.246 \\
Export textured mesh & 0.845 & 0.325 \\
\midrule
Total TextureGen critical path 
& \cellcolor{assetgenfullrow}\textbf{18.01} 
& \cellcolor{assetgennanorow}\textbf{6.62} \\
\bottomrule
\end{tabular}
\end{table}

\subsection{Kernel and Precision Optimization}%
\label{sec:kernel-precision}

We apply three optimizations to reduce latency.
First, FlashAttention 3~\citep{fa3} is used in both diffusion transformers, reducing latency by 20\% with no observed quality change.
Second, non-blocking host-to-device transfers overlap data copies with kernel execution, saving an additional 11\%.
Third, default-mode graph compilation~\citep{torch_compile} is applied to the diffusion transformers and backprojection kernels, fusing element-wise operations and reducing kernel-launch overhead.

We also reduce numerical precision where possible, applying FP8 and INT8 quantization in both diffusion transformers, with precision assignments determined by sensitivity analysis.
For TextureGen, embedding layers, the first and last transformer blocks, and output projections remain in higher precision, as these components are visibly sensitive to quantization.

\subsection{Pipeline Optimization}%
\label{sec:pipeline-parallel}

We can hide most of the latency of other steps
by running them in parallel.
We parallelize independent work within a request.

Within each stage, we identify operations that have no data or resource dependency and schedule them concurrently.
In TextureGen, several expensive preprocessing tasks depend only on the processed mesh and camera layout, not on the generated colors.
These include UV-to-image sampling maps, visibility and incidence masks, anisotropic filter parameters, and KD-tree neighbor structures for inpainting.
Because the processed mesh is fixed before texture diffusion begins, these geometry-dependent computations run in a separate worker process while diffusion sampling runs on the GPU\@.

Irreducible dependencies increase latency.
In our design, TextureGen needs to wait for mesh simplification and normal baking to be complete, adding to the critical path.

MeshGen and TextureGen are implemented as separate services.
MeshGen runs on one H100 GPU and writes the simplified mesh, UV layout, baked normal map, and associated metadata to shared storage, which is read by TextureGen to generate the texture map.
For \assetgenpro, TextureGen uses two H100 GPUs to split the conditional and unconditional CFG branches, whereas \assetgenflash uses only one as CFG is distilled.
The data transfer cost adds roughly 1--2 seconds compared to an ideal in-process handoff, and this overhead is included in the reported results.

\subsection{Latency Results}%
\label{sec:latency_results}

We report latency in three ways.
First, \cref{tab:latency-breakdown-meshgen,tab:latency-breakdown-texturegen} break down the major MeshGen, geometry processing, and TextureGen stages.
These values are averaged over 100 benchmark images and are intended to show where time is spent in the system.
These timings should not be summed to obtain the total system latency as many of these steps run in parallel.
Conversely, these timings do not account for the cost of synchronization, tensor handoff, worker scheduling, service-to-service transfer, and serialization overheads that are not always assigned to a single row.
The ``Total'' rows report measured module-level critical-path latency (not the numerical sum of the rows), while the final end-to-end numbers are measured directly from input image to exported asset.

Second, \cref{tab:e2e-latency} reports the time required to serve a user request end-to-end, including an additional 0.65 seconds required to segment out the object in the user-provided image.
These numbers account for details of our hosted infrastructure and therefore should be interpreted as representative user-visible latency, not as a hardware-normalized benchmark.

\begin{table}[htbp]
\centering
\caption{End-to-end wall-clock latency comparison. AssetGen measurements capture the complete pipeline from input image to exported textured mesh, averaged across 100 benchmark assets, and include a shared 0.65s image-segmentation stage. Baseline latencies reflect the time from web submission to result availability averaged over five trials, excluding export and download overhead. Because these baselines operate on unknown hosted infrastructure, the values serve as operational latency references rather than hardware-normalized benchmarks.
}%
\vspace{-3mm}
\label{tab:e2e-latency}
\small
\setlength{\tabcolsep}{4pt}
\begin{tabularx}{\linewidth}{@{}l *{6}{>{\centering\arraybackslash}X}@{}}
\toprule
\textbf{Method}
& \textbf{AssetGen}
& \textbf{AssetGen Flash}
& \textbf{\methodA}
& \textbf{\methodB}
& \textbf{\methodC}
& \textbf{\methodD}\\
\midrule
\textbf{Wall-clock latency}
& $\sim$30s
& $\sim$14s
& $\sim$180s
& $\sim$125s
& $\sim$100s
& $\sim$130s \\
\bottomrule
\end{tabularx}
\end{table}

\begin{table}[t]
\centering
\caption{Ablation study of latency improvements evaluated on a representative test asset. Each row reports the latency after cumulatively applying the corresponding optimization.}
\vspace{-3mm}
\label{tab:latency_ladder}
\footnotesize
\setlength{\tabcolsep}{8pt}
\renewcommand{\arraystretch}{1.15}
\begin{tabular}{@{}l r  l r@{}}
\toprule
\multicolumn{2}{c}{\textbf{MeshGen + Geometry}} & \multicolumn{2}{c}{\textbf{TextureGen}} \\
\midrule
Stage & Lat.\,(s) & Stage & Lat.\,(s) \\
\midrule
Baseline                 & 103.15 & Baseline                 & 117.68 \\
+ CuMesh                 & 78.15  & + Prog.\ Distillation    & 41.42  \\
+ Prog.\ Distillation    & 53.10  & + FA3 + compile          & 31.21  \\
+ FA3 + compile          & 32.79  & + Precompute backproject data & 27.11  \\
+ I/O optimization       & 27.97  & + Rendering optimization & 23.23  \\
+ Parallel UV unwrap     & 21.09  & + Dual-GPU CFG branch split &  \textbf{17.85} \\
+ DRTK normal baking     & \textbf{12.25} &  & \\
\bottomrule
\end{tabular}
\end{table}

For AssetGen, latency is measured from input image to exported textured mesh and includes image segmentation, service orchestration, serialization, and asset export.
For commercial baselines, we measure website wall-clock time over five generations, from submitting the image to the point where the generated result becomes available in the web interface; we do not include additional export or download time.
Because these systems run on unknown hosted infrastructure and may include website-side scheduling or queueing, these measurements should be interpreted as operational latency references rather than controlled hardware-normalized comparisons.
In our measurements, hosted high-quality image-to-3D systems operate around the near-minute to two-minute regime, while \assetgenpro achieves similar quality in just 30 seconds, and \assetgenflash generates a preview in less than 15.

Third, \cref{tab:latency_ladder} provides a ladder analysis of our acceleration techniques.
Unlike \cref{tab:latency-breakdown-meshgen,tab:latency-breakdown-texturegen}, which average timings over 100 benchmark images, here we evaluate variants of the system on one fixed input and measure the cumulative effects of the optimizations.
For MeshGen, replacing CPU-bound geometry processing and then applying progressive distillation, Flash Attention 3, graph compilation, I/O overlap, parallel UV segmentation, and DRTK normal baking reduces the fixed-sample latency from 103.15s to 12.25s.
For TextureGen, progressive distillation provides the largest single reduction, while kernel optimizations, precomputed backprojection data, rendering optimizations, and the two-GPU CFG branch split further reduce the default quality path.

\section{Evaluation}%
\label{sec:evaluation}

This section evaluates the quality of the assets produced by \method against leading image-to-3D generators.
We introduce two benchmarks assessing generation quality for general objects (\cref{sec:assetbench}) and for characters (\cref{ssec:pta}), respectively.
We further conduct blind human evaluation in \cref{ssec:human_eval}.

\paragraph{Baselines.}

We compare \assetgenpro and \assetgenflash against \methodA, \methodB, \methodC, and \methodD.
We pass the same reference image to each system and obtain the corresponding assets. We use a target polygon count of 23K.

\subsection{AssetBench: General Object Generation}%
\label{sec:assetbench}

To assess the quality of generated assets across diverse categories, we introduce AssetBench, a benchmark of 101 high-quality 3D assets manually curated and reviewed by technical artists.
The dataset spans diverse categories including vehicles, daily-use objects, animals, and some characters.
For each sample, we render an isometric/frontal reference view of the ground-truth 3D asset, which serves as the image prompt for all evaluated methods.

We evaluate image-to-3D generation along two axes.
Shape quality is measured using \emph{Chamfer Distance (CD)}, the symmetric Chamfer Distance between uniformly sampled surface points after unit-cube normalization and ICP alignment over 24 canonical orientations with gradient-based refinement, and \emph{Volumetric IoU}, the occupancy IoU estimated from $10^5$ uniformly sampled points in $[-0.5,0.5]^3$ after the same ICP alignment.
Faithfulness to the image prompt is measured using \emph{VLM-as-a-Judge}, where a VLM extracts attribute-value pairs from the reference image, converts them into binary verification questions, and answers each question for 9 rendered views; we score \emph{yes}/\emph{partially}/\emph{no} as $1/0.5/0$ and average over questions and views~\citep{duggal2025eval3d,hu2023tifa}, and \emph{CLIP Similarity}, the mean CLIP ViT-L/14 cosine similarity between generated and ground-truth renders from 9 uniformly spaced rendered views of the generated mesh.

\begin{table}[htbp]
\centering
\caption{Quantitative results on AssetBench. The best values are highlighted in \textbf{bold} and the second-best are \underline{underlined}. Arrows indicate whether higher ($\uparrow$) or lower ($\downarrow$) values are optimal.}%
\label{tab:assetbench_results}
\scalebox{1.0}{
\setlength{\tabcolsep}{8pt}
\small
\begin{tabular}{l cc cc}
\toprule
& \multicolumn{2}{c}{\textit{Shape Quality}} & \multicolumn{2}{c}{\textit{Fidelity}} \\
\cmidrule(lr){2-3} \cmidrule(lr){4-5}
\textbf{Method} & IoU\,$\uparrow$ & CD\,$\downarrow$ & VLM\,$\uparrow$ & CLIP\,$\uparrow$ \\
\midrule
AssetGen  & 0.4985 & \underline{0.0098} & \textbf{0.7048} & {\textbf{0.8993}} \\
AssetGen Flash & \underline{0.4997} & \underline{0.0098} & \underline{0.6977} & \underline{0.8978} \\
\midrule
\methodA          & 0.4993 & 0.0102 & 0.6921 & 0.8826 \\
\methodB & \textbf{0.5298} & \textbf{0.0087} & 0.6500 & 0.8610 \\
\methodC & 0.4522 & 0.0103 & 0.6482 & 0.8528 \\
\methodD          & 0.4849 & 0.0119 & 0.6905 & 0.8869 \\
\bottomrule
\end{tabular}
}
\end{table}

\paragraph{Results.}
\Cref{tab:assetbench_results} shows a clear split between geometry and appearance/reference fidelity.
\methodB~leads the volumetric geometry metrics, with the best IoU and Chamfer Distance, indicating the strongest geometric reconstruction among the evaluated baselines.
AssetGen is close on geometry while leading the reference-fidelity metrics, achieving the highest VLM verification score and GT-CLIP similarity.
This suggests that AssetGen's main advantage is not a uniformly stronger geometry prior, but a stronger end-to-end asset pipeline for preserving reference appearance and producing textured assets at much lower latency.

\begin{figure*}[t]
\centering
\includegraphics[width=\textwidth]{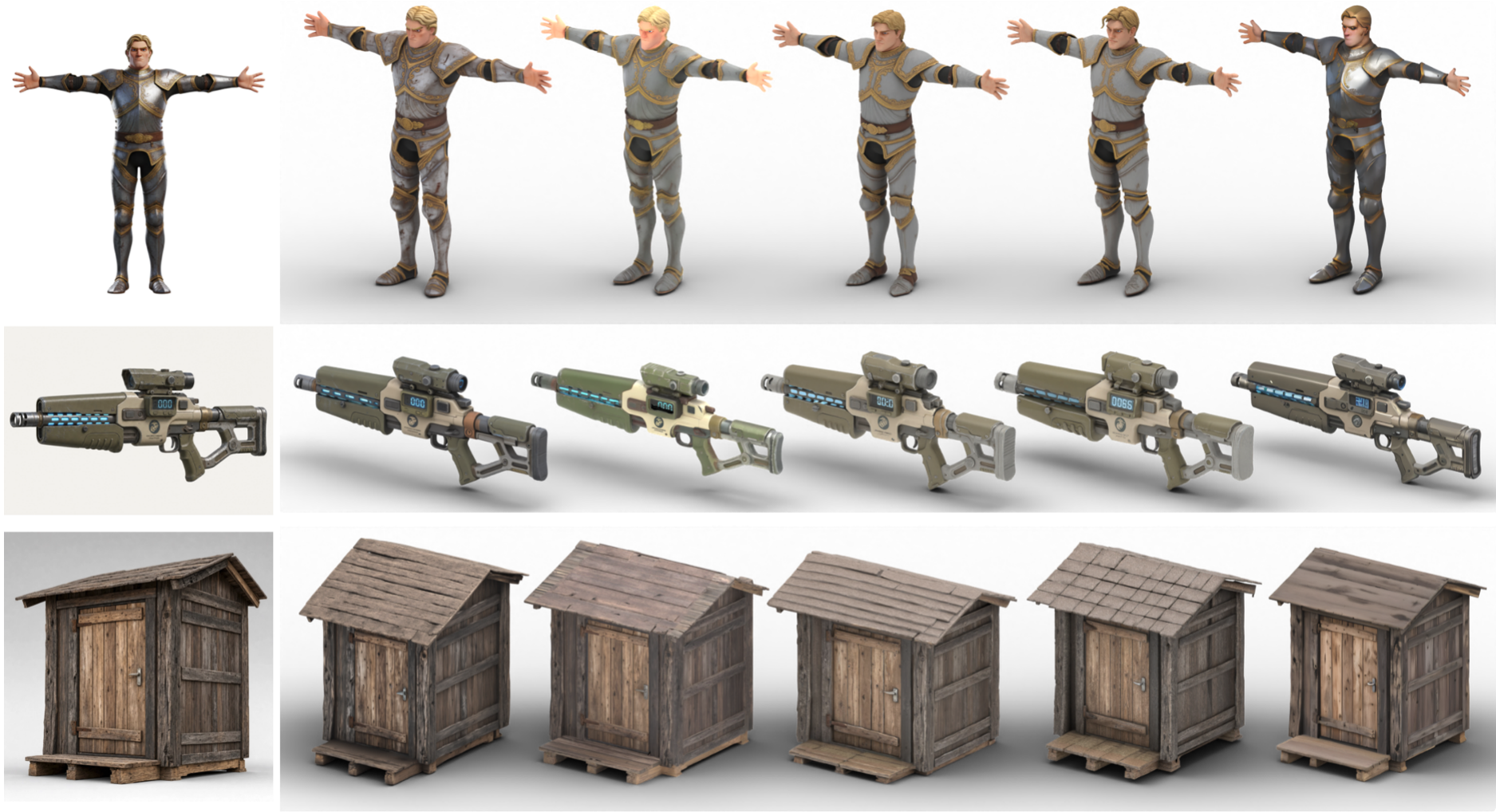} \\
\begin{tabular}{@{}p{0.16\textwidth}@{\hskip 18pt}p{0.16\textwidth}@{\hskip 0pt}p{0.16\textwidth}@{\hskip 0pt}p{0.16\textwidth}@{\hskip 0pt}p{0.16\textwidth}@{\hskip 0pt}p{0.16\textwidth}@{}}
\centering Input Image & \centering AssetGen &  \centering\methodA & \centering\methodB & \centering \methodC &  \centering\methodD \\
\end{tabular}
\caption{Qualitative comparison against commercial baselines. From left to right: the input image, AssetGen, \methodA, \methodB, \methodC, and \methodD.}
\label{fig:baseline}
\end{figure*}

\begin{figure*}[t]
\centering
\includegraphics[width=\textwidth]{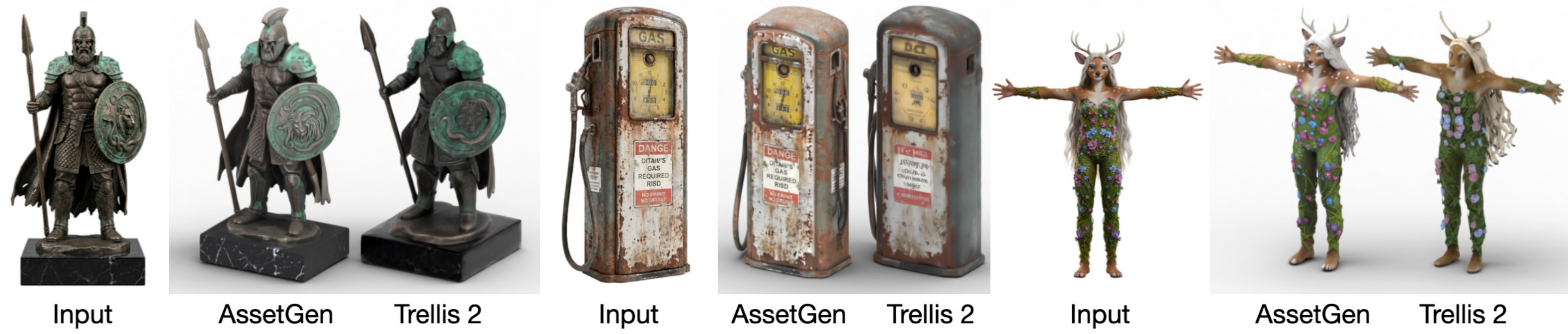}
\caption{Qualitative comparison between AssetGen and Trellis~2. AssetGen preserves fine details more faithfully, as shown by the legible texts on the gas pump (``GAS'' ``DANGE'' ) and intricate details of the character's body and facial components.}
\label{fig:baseline_trellis}
\end{figure*}

\begin{figure*}[ht]
\centering
\includegraphics[width=\textwidth]{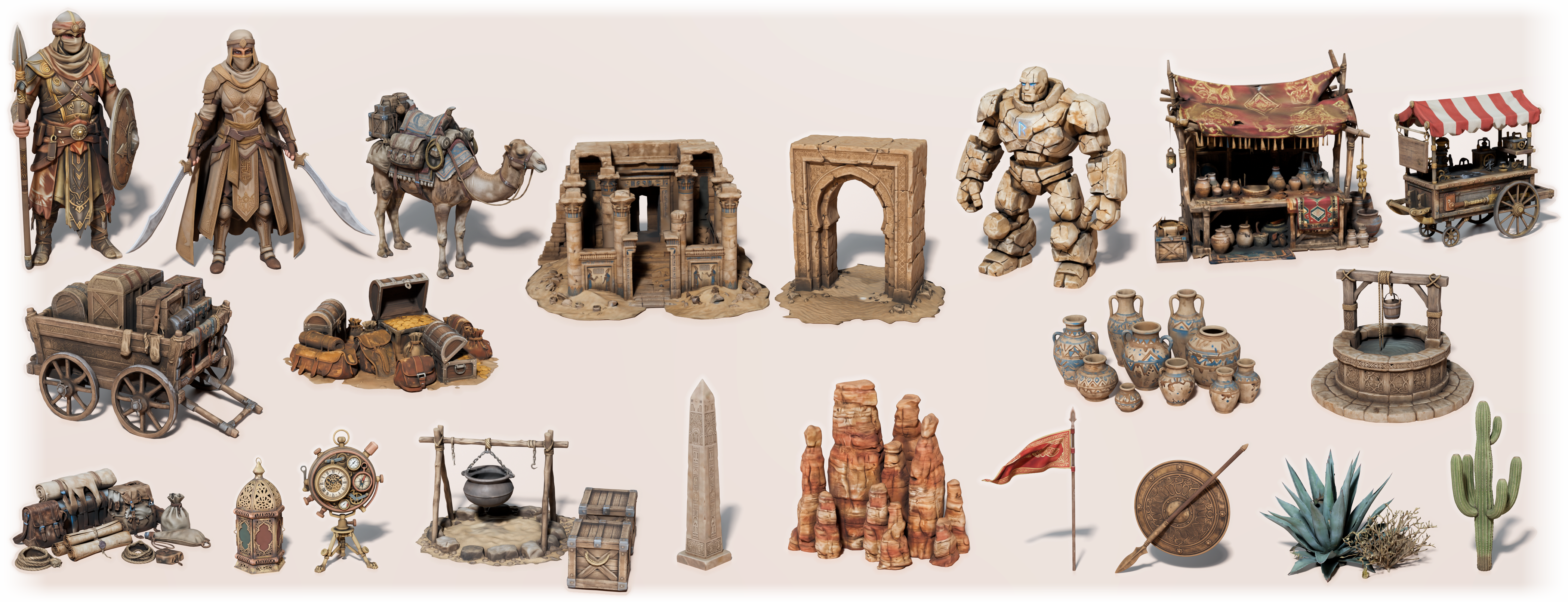}
\caption{AssetGen qualitative gallery featuring sand-themed assets. This default configuration utilizes the two-stage MeshGen pipeline and high-fidelity TextureGen to prioritize visual quality, delivering production-ready assets in approximately 30 seconds end-to-end.}%
\label{fig:texturegen_gallery}
\end{figure*}

\begin{figure*}[ht]
  \centering
    \includegraphics[width=\textwidth]{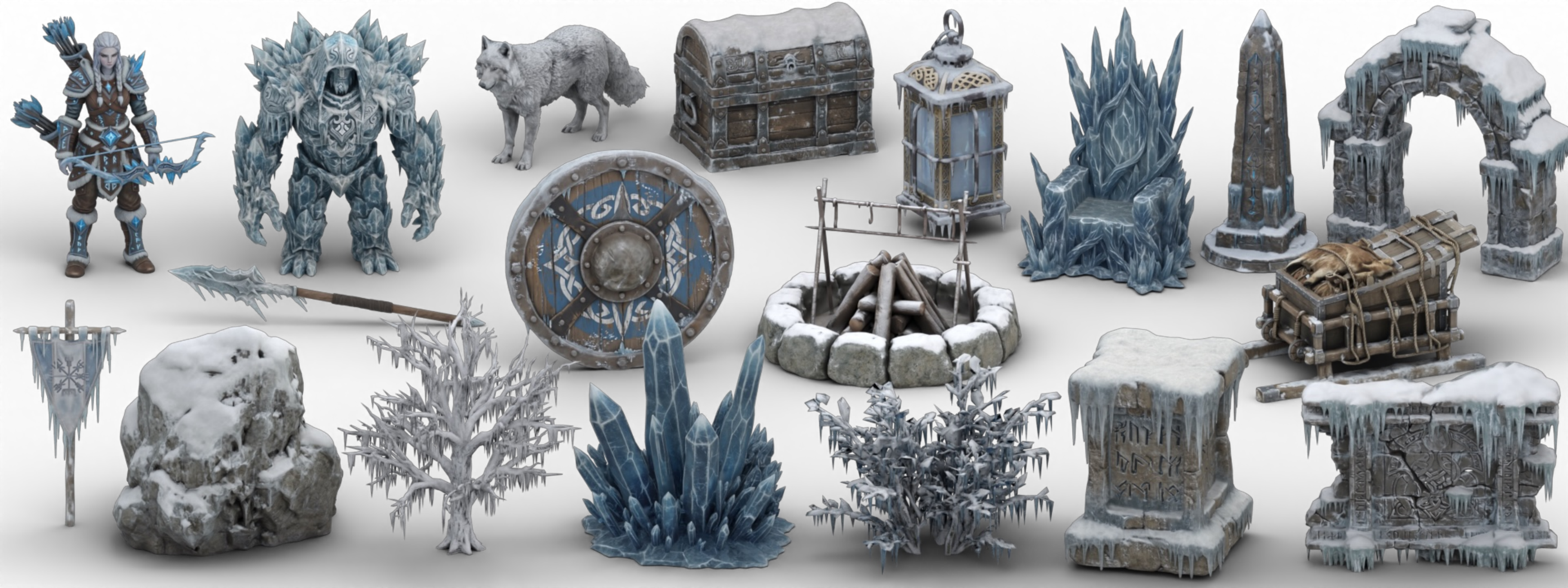}
  \caption{AssetGen qualitative gallery featuring ice-themed assets. }
  \label{fig:gallery_assetgen}
\end{figure*}

\subsection{CharacterBench: Character Generation}%
\label{ssec:pta}

CharacterBench is a benchmark of 100 character reference images designed to evaluate image-to-3D generation for characters.
Unlike AssetBench, no ground-truth meshes are provided.
Instead, the evaluation focuses on perceptual quality dimensions uniquely challenging for character generation---facial likeness, anatomical hand correctness, and full-body fidelity to the reference---using specialized VLM-based and CLIP-based protocols.

\begin{table}[htbp]
\centering
\caption{Quantitative results for character generation on CharacterBench. The best value in each column is highlighted in \textbf{bold} and the second-best are \underline{underlined}. All metrics favor higher values ($\uparrow$). }%
\label{tab:pta_results}
\footnotesize
\scalebox{0.95}{
\begin{tabular}{l cc cc c cc}
\toprule
& \multicolumn{2}{c}{\textit{Face Quality}}
& \multicolumn{2}{c}{\textit{Face Fidelity}}
& \textit{Hand}
& \multicolumn{2}{c}{\textit{Overall Fidelity}} \\
\cmidrule(lr){2-3}
\cmidrule(lr){4-5}
\cmidrule(lr){6-6}
\cmidrule(lr){7-8}
\textbf{Method}
& Frontal & Profile
& Face CLIP & Face Ref
& Finger
& Ref Fid & CLIP \\
\midrule
AssetGen
& \textbf{0.748} & 0.196
& \textbf{0.838} & \textbf{0.893}
& 0.784
& \textbf{0.903} & \underline{0.777} \\
AssetGen Flash
& 0.554 & 0.191
& 0.822 & 0.888
& 0.773
& \underline{0.901} & \textbf{0.780} \\
\midrule
\methodA
& 0.558 & 0.219
& 0.812 & 0.882
& 0.732
& 0.895 & 0.772 \\
\methodB
& \underline{0.695} & \textbf{0.230}
& \underline{0.832} & \underline{0.890}
& \textbf{0.948}
& \underline{0.901} & 0.767 \\
\methodC
& 0.590 & \textbf{0.230}
& 0.814 & 0.879
& \underline{0.928}
& 0.891 & 0.761 \\
\bottomrule
\end{tabular}
}
\end{table}

\paragraph{Metrics.}

We evaluate along four character-specific axes.
First, we assess the quality of the generated face.
We first select a VLM-verified camera distance that keeps the full face visible, then evaluate frontal and profile renders of the face.
\emph{Frontal Face Score} uses a VLM to check frontal renders for symmetry, natural expression, correct eye count and placement, mouth-nose alignment, and absence of blurred facial regions.
A separate eye-quality rubric evaluates each eye for asymmetry and blur.
The final score averages the normalized QA score and the eye score.
\emph{Profile Face Score} evaluates left and right profile renders for texture artifacts, ear visibility, and clean face/non-face boundaries.
Scores are averaged across both profiles, and eye evaluation is omitted because profile occlusion makes it unreliable.

Next, we evaluate face fidelity relative to the reference identity.
\emph{Face CLIP Similarity} segments the reference face with SAM3~\citep{carion2025sam}, crops rendered frontal faces from three views using per-pixel face-index maps, and averages CLIP ViT-L/14 cosine similarity between the reference crop and rendered crops.
\emph{Face Reference Fidelity} uses VLM-generated verification questions to assess facial attributes such as features, colors, proportions, and markings.
The VLM verifies these questions on rendered views.

We also evaluate the quality of the generated \emph{hands}.
\emph{Finger Score} renders each hand from multiple views, segments candidate regions, and uses a VLM to classify each region as finger or non-finger.
View counts are aggregated by majority vote when the palm faces the camera, and by maximum count otherwise to account for occlusion.
The score penalizes missing and hallucinated fingers symmetrically.

Finally, we evaluate full-body correspondence to the input reference.
\emph{Reference Fidelity} applies the same verification-question protocol to clothing, proportions, accessories, materials, and markings, reporting the fraction of correctly answered questions.
Questions are generated by a VLM\@.
\emph{CLIP Image Similarity} computes the mean CLIP ViT-L/14 cosine similarity between the input reference image and 24 rendered views, excluding top and bottom views.

\paragraph{Results.}
\Cref{tab:pta_results} reports character-specific metrics on CharacterBench.
AssetGen achieves the best frontal face quality (0.748) and leads both face-fidelity metrics, Face CLIP (0.838) and Face Ref (0.893), indicating strong preservation of the reference identity.
AssetGen also obtains the highest overall reference-fidelity score (0.903), reflecting better consistency in clothing, accessories, proportions, and markings.
The clearest remaining gap is hand structure: \methodB~achieves the highest finger score, showing a stronger hand-geometry prior.
AssetGen (Flash) remains close to the default configuration on reference-fidelity metrics and slightly improves full-body CLIP similarity, consistent with its role as a faster configuration that preserves usable visual quality.

\begin{figure*}[t]
\centering
\includegraphics[width=\textwidth]{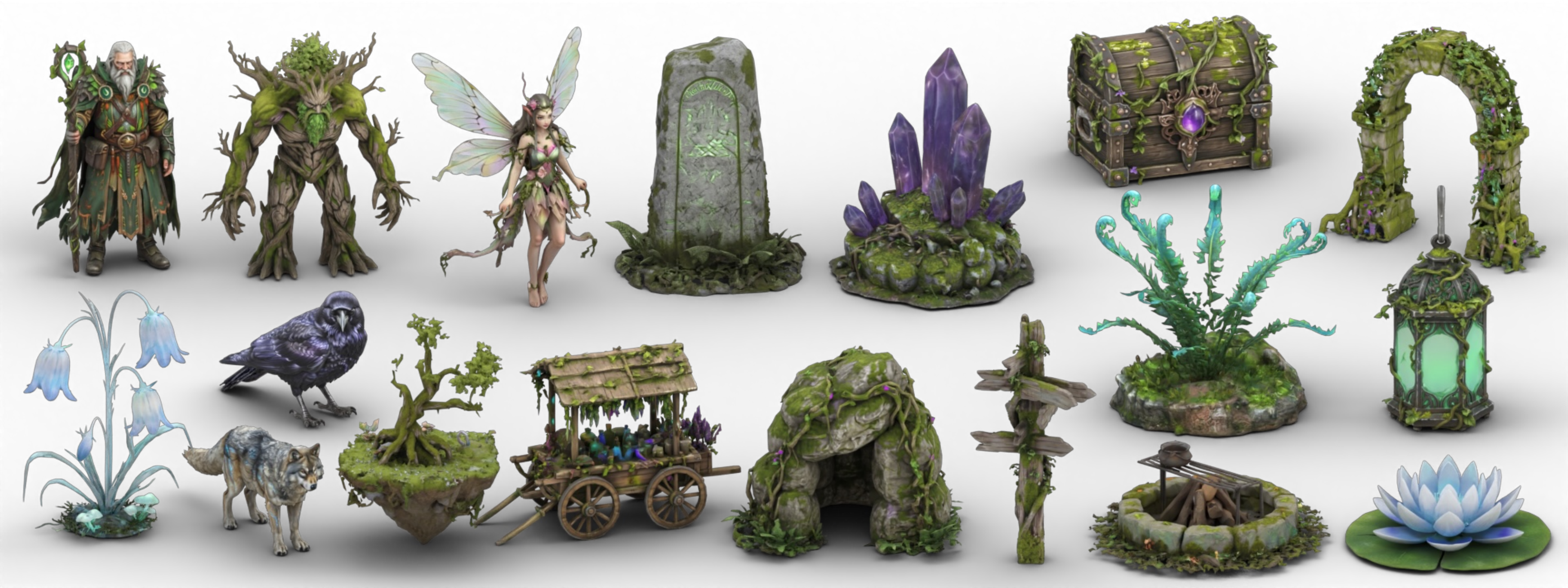}
\caption{AssetGen (Flash) qualitative gallery featuring magic forest assets.}
\label{fig:gallery_assetgen_flash}
\end{figure*}

\subsection{Human Evaluation}%
\label{ssec:human_eval}

We further conduct a blind human evaluation on 195 image-conditioned assets spanning general objects, characters, buildings, and other common asset categories.
To reduce bias from benchmark construction, the input images were collected by artists, who were asked to provide diverse concept images they considered useful for 3D asset creation.
We did not select or inspect these images during benchmark construction.
Evaluators inspect each generated result in an interactive 3D viewer with access to the final mesh, texture, UV layout, and wireframe.
We do not use turntable videos or fixed rendered views for this study; the goal is to evaluate the asset itself under direct inspection.
Method identity is hidden from annotators.

Each asset is scored on an absolute 1--5 scale, where 1 indicates a severe failure and 5 indicates an excellent result.
We group the rubric into three categories.
\emph{General quality} averages prompt adherence, style adherence, usability, and aesthetic quality.
\emph{Geometry quality} averages distortion, proportion, contiguity, accuracy, scale, alignment, topology, and UV layout.
Geometry is evaluated on the final simplified mesh for all methods, matching the representation that would be used downstream.
\emph{Texture quality} averages consistency, accuracy of materials and shapes, clarity of details and patterns, and visible artifacts.
Category scores are computed as the mean over their respective sub-criteria.

\begin{table}[ht]
\centering
\caption{Blind human evaluation across 195 assets. All scores are reported on an absolute 1--5 scale ($\uparrow$ higher is better). The general, geometry, and texture ratings represent averages over the specific subcriteria defined in \cref{ssec:human_eval}. Note that geometry quality is evaluated directly on each method's final simplified mesh.}
\label{tab:human_eval}
\small
\begin{tabular}{lccc}
\toprule
\textbf{Method} & \textbf{General} $\uparrow$ & \textbf{Geometry} $\uparrow$ & \textbf{Texture} $\uparrow$ \\
\midrule
{AssetGen} & \textbf{2.80} & \textbf{2.96} & \textbf{2.68} \\
{AssetGen Flash} & {2.63} & {2.88} & {2.55} \\
{\methodA} & {2.61} & {2.86} & {2.58} \\
\bottomrule
\end{tabular}
\end{table}

\Cref{tab:human_eval} shows that AssetGen receives the highest average score in all three categories.
The margins are modest rather than overwhelming, which is the right interpretation for a strong commercial baseline: the main result is that the default AssetGen path reaches comparable-to-better inspected asset quality while operating at much lower end-to-end latency.
Flash remains close to \methodA~on general and geometry quality while trailing the full AssetGen model, reflecting the intended quality-responsiveness trade-off.

\section{Qualitative Results}
\label{sec:qualitative_results}

In this section, we present qualitative results that complement the quantitative evaluation. 
We include matched comparisons with commercial systems, a comparison with Trellis 2~\citep{xiang2025native}, galleries from the default AssetGen configuration and the faster AssetGen (Flash) configuration, and examples on real-world images.

\paragraph{Qualitative comparison.}
\cref{fig:baseline} compares AssetGen against \methodA, \methodB, \methodC, and \methodD~under matched input and rendering conditions. 
All methods are evaluated without PBR materials. 
Overall, the commercial baselines are strong and produce compelling assets. 
\methodB~often produces the strongest geometry, consistent with our quantitative results.
AssetGen produces competitive geometry while generally showing cleaner and richer textures, and stronger reference-image fidelity.

\paragraph{Comparison with Trellis 2.}
\cref{fig:baseline_trellis} compares AssetGen with Trellis 2, a strong open-source image-to-3D baseline.
We use the Trellis 2 output as recommended and keep its final mesh at one million polygons.
This is a much denser representation than AssetGen, whose outputs are simplified to about 23K.
At the Trellis 2 scale, UV unwrapping alone can take close to a minute.
Qualitatively, Trellis 2 is less robust than AssetGen and the commercial baselines on complex inputs.
For example, in the second example, Trellis 2 fails to recover the letters cleanly; and in the third example, the fingers are distorted despite the high polygon count.
\paragraph{AssetGen gallery.}
\cref{fig:texturegen_gallery} and \cref{fig:gallery_assetgen} show generations from the default AssetGen configuration.
Both the input images and the 3D meshes are single-shot generations without cherry picking.

\paragraph{AssetGen (Flash) gallery.}
\cref{fig:gallery_assetgen_flash} shows results from AssetGen (Flash).
Both the input images and the 3D meshes are single-shot generations without cherry picking.

\begin{figure}[t]
    \centering

    \includegraphics[width=0.85\columnwidth]{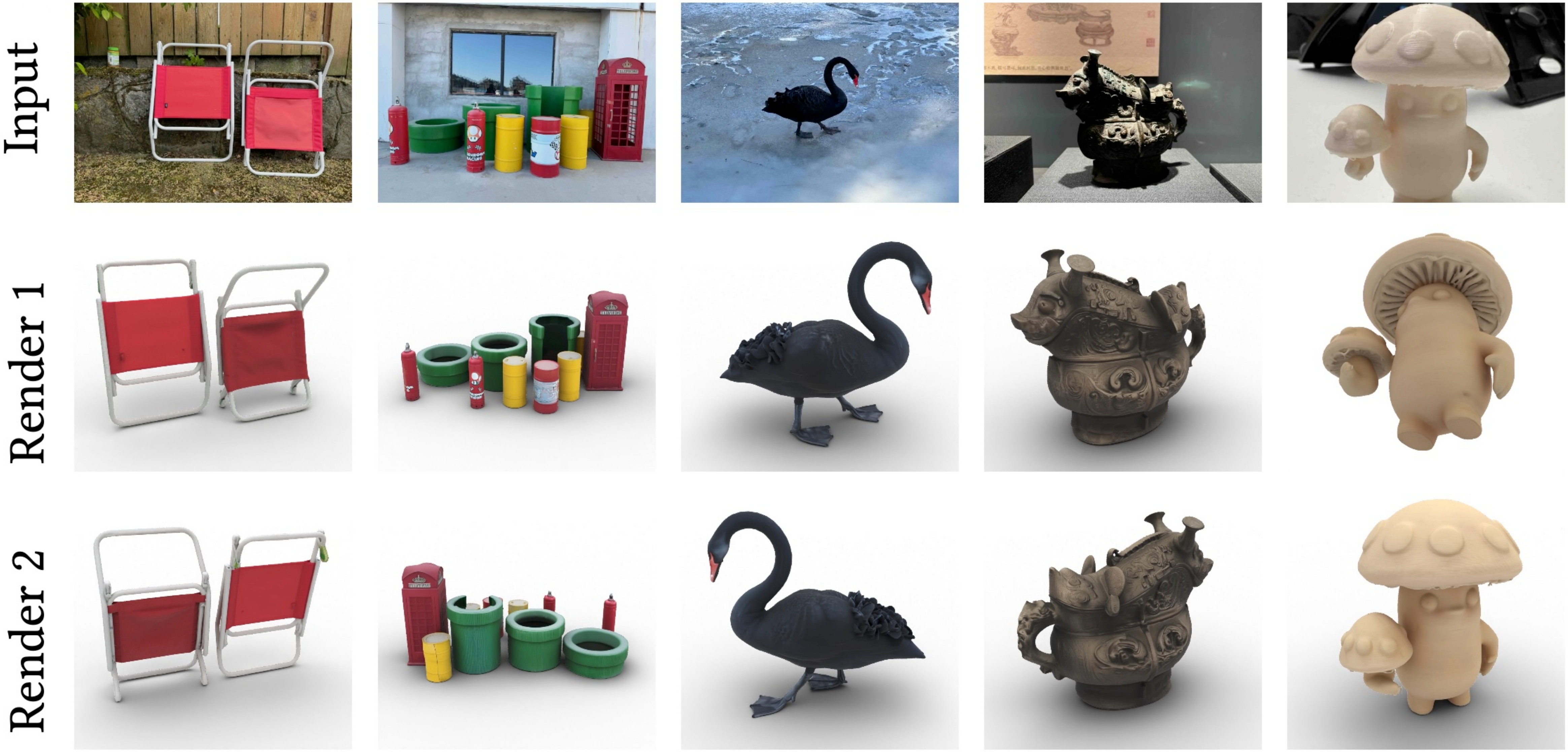}

    \caption{Asset generation from real-world images. Each column displays an in-the-wild input photograph alongside two rendered views of the resulting AssetGen mesh. These examples demonstrate the pipeline's robustness to natural lighting, background clutter, and partial occlusion.}
    \label{fig:real_world_reconstruction}
\end{figure}

\paragraph{Real-World Inputs}
\label{sec:real_world_inputs}

\cref{fig:real_world_reconstruction} shows qualitative examples on real-world photographs.
These examples suggest that the same image-to-asset pipeline can also produce plausible textured meshes from real-world inputs, even though the model is trained primarily on synthetic 3D data.
For example, in the second column, AssetGen recovers a complete bucket shape despite partial occlusion, illustrating how its learned shape prior can plausibly complete unseen regions.

\section{Conclusion and Limitations}
\label{sec:conclusion}

\method provides a practical path from a single reference image to an
explicit textured 3D asset. The default AssetGen configuration produces
a simplified, UV-unwrapped, normal-baked textured mesh in approximately
30 seconds on H100 GPUs. AssetGen (Flash) runs in approximately 14 seconds in our measured deployment, providing a faster operating point for rapid exploration, preview, and agentic workflows. 
Through extensive quantitative evaluation, qualitative comparison, and blind human inspection, we show that AssetGen is competitive in asset quality with leading commercial image-to-3D systems while substantially reducing user-facing generation time.

While our results demonstrate a fast and deployable asset generation pipeline, several limitations remain:

\emph{Single-image ambiguity.}
Single-image 3D generation is inherently underconstrained. The model
must infer occluded surfaces, which can fail for rare categories,
unusual viewpoints, asymmetric objects, or complex back-side structure.
Symmetric objects are generally more reliable; irregular chairs,
sculptures, and objects with hidden mechanical parts remain more
challenging.

\emph{Topology constraints.}
Marching-cubes extraction produces arbitrary triangle meshes. The
simplification pipeline preserves surface fidelity and keeps assets
lightweight, but it does not enforce clean topology,
artist-friendly edge flow, skeletal rigging constraints, or
animation-ready deformation behavior. Assets intended for rigged
characters may still require retopology.

\emph{Rigging and animation.}
Static textured meshes are only one step toward interactive characters.
Integrating auto-rigging, skin-weight prediction, blend-shape
generation, and motion synthesis would close the gap between generated
assets and animated avatars, but these tasks require topology-aware
reasoning and scarce rigged training data.

\clearpage
\newpage
\bibliographystyle{assets/plainnat}
\bibliography{ref_fixed}

\end{document}